\documentclass[twocolumn,showpacs,preprintnumbers,pra,amsmath,amssymb,superscriptaddress]{revtex4}
\usepackage{bm}
\newcommand{\bfsfI}{\mbox{\sffamily\bfseries{I}}}
\newcommand{\bfsfG}{\mbox{\sffamily\bfseries{G}}}
\newcommand{\bfsfU}{\mbox{\sffamily\bfseries{U}}}
\newcommand{\bfsfc}{\mbox{\sffamily\bfseries{c}}}
\newcommand{\bfsff}{\mbox{\sffamily\bfseries{f}}}
\newcommand{\half}{{\textstyle \frac{1}{2}}}
\def\hbarit {{\mathchar'26\mkern-11muh}} 

\begin{document}
\title{Field quantization in inhomogeneous absorptive dielectrics}
\date{\today
}
\author{L.G.~Suttorp}
\affiliation{Institute for Theoretical Physics, University of
Amsterdam, Valckenierstraat 65, NL-1018 XE  Amsterdam, The
Netherlands}
\author{Martijn Wubs} 
\affiliation{Complex Photonic Systems, Faculty of Science and Technology,
University of Twente, P.O. Box 217, NL-7500~AE~~Enschede, The Netherlands}
\affiliation{Van der Waals-Zeeman Institute, University of Amsterdam,
Valckenierstraat 65, NL-1018 XE Amsterdam, The Netherlands}

\begin{abstract}
The quantization of the electromagnetic field in a three-dimensional
inhomogeneous dielectric medium with losses is carried out in the framework
of a damped-polariton model with an arbitrary spatial dependence of its
parameters. The equations of motion for the canonical variables are solved
explicitly by means of Laplace transformations for both positive and
negative time. The dielectric susceptibility and the quantum noise-current
density are identified in terms of the dynamical variables and parameters
of the model. The operators that diagonalize the Hamiltonian are found as
linear combinations of the canonical variables, with coefficients depending
on the electric susceptibility and the dielectric Green function. The
complete time dependence of the electromagnetic field and of the dielectric
polarization is determined. Our results provide a microscopic justification
of the phenomenological quantization scheme for the electromagnetic field
in inhomogeneous dielectrics.
\end{abstract}
\maketitle

\section{Introduction\label{sectionintro}}

The quantization of the electromagnetic field in a linear dielectric medium
has been the subject of many investigations since the first treatment by
Jauch and Watson \cite{JW48}. In a homogeneous and non-dispersive medium
the photon is associated with the transverse part of the field, which can
neatly be distinguished from its longitudinal part. In contrast, in an
inhomogeneous non-dispersive medium the transverse and the longitudinal
degrees of freedom get coupled, which renders the generalization of the
quantization scheme to that case less straightforward. However, the
quantization can still be accomplished by employing a generalized
transverse gauge, which depends on the dielectric constant
\cite{KVW87,GL91,KW92,DGK96,DB97,WSL03}.

For a dielectric medium with dispersion the quantization procedure
described above runs into problems. In fact, since dispersion in a
dielectric medium is inextricably connected to extinction, one should
account for the effect of losses in the quantization procedure. Huttner and
Barnett \cite{HB92a,HB92b} were the first to use the Hopfield polariton
model \cite{H58} for the description of a dielectric with losses. To
incorporate losses they coupled the dielectric polarization to a bath of
oscillators, which causes a damping of the polaritons. Subsequently, upon
assuming the medium to be homogeneous, they were able to diagonalize the
Hamiltonian of this damped-polariton system, and to establish explicit
formulas for the electromagnetic field and the dielectric polarization in
terms of the diagonalizing operators. As a direct application of the model
they evaluated the change in the atomic decay due to a dielectric
environment \cite{BHL92}. Later on, their work has been reformulated and
extended in various ways. In \cite{DF98a} and \cite{DF98b} a simplified
expression for the dielectric constant of the model was found. An
alternative description of the model in terms of path integrals was given
in \cite{B99}. In \cite{WS01} Laplace transformations were employed to
simplify the diagonalization process considerably. Finally, in \cite{JRH03}
these transforms were used to formulate the equations governing the
dynamics of the inhomogeneous model, although a complete diagonalization
was not attempted in that case.

As stated above, the treatment in \cite{HB92a,HB92b} is confined to the
homogeneous damped-polariton model. For such systems, a systematic use of
spatial Fourier transforms greatly helps in carrying out the
diagonalization of the Hamiltonian. Somewhat later it was realized
\cite{GW95,MLBJ95,GW96a,ML96,GW96b,DKW98,SKW98} that the quantization of
the fields in inhomogeneous media could be achieved in position space by
adding a noise term to the Maxwell equations in a phenomenological way. In
agreement with the fluctuation-dissipation theorem, one then postulates
suitable commutation relations for this noise term. Accounting for the
Kramers-Kronig relations of the dielectric media is found to be essential
in defining field operators with standard commutation relations. An
alternative formulation \cite{T98} of the quantization procedure by means
of auxiliary fields has been shown to be completely equivalent \cite{T01}.

The phenomenological quantization scheme has been extended to magnetic and to
anisotropic media, and to finite media with gain (see the reviews \cite{KSW01,
LP02} and references therein). The scheme has been applied to atomic decay
\cite{BHLM96,SKWB99,SKW99a,DKW00}, to energy transfer \cite{DKW02a}, and to
resonant dipole interactions \cite{DKW02b}. The phenomenological scheme has also
been used to study the properties of electromagnetic field operators in systems
with optical cavities or beam splitters \cite{BGHI96,BJGL98}, and with
dielectric slabs \cite{SSG00}. For these optical components input-output
relations have been derived. It was found that extinction usually has adverse
effects on nonclassical properties of light, such as squeezing, nonclassical
correlations and entanglement \cite{AL99,SKOW00}.

Although the phenomenological quantization of the electromagnetic field in
absorptive dielectrics has been very successful, its connection to the
damped-polariton model has been established only for the special case of a
homogeneous dielectric medium. In fact, one would like to see whether the
crucial properties of the noise term, which are postulated in the
phenomenological approach, could be derived from the polariton model in the
general inhomogeneous case, as has been noted on various occasions
\cite{DKW02a,DKW02b,YG96}. Since the model furnishes a precise Hamiltonian
description of the interaction between the damped dielectric and the
electromagnetic field, one would have obtained in this way a microscopic
justification of the phenomenological quantization scheme. To arrive at
this goal, one has to express the noise term in the canonical variables of
the model, so that its properties can be determined. Once the noise term
has been found, it can be used for the complete diagonalization of the
Hamiltonian. In the following we shall show how this can be achieved by
employing a Laplace-transform technique as in \cite{WS01,JRH03}.

The paper is organized as follows. In Sec.\ \ref{sectioneqmotion} the model
is defined and the equations of motion for the canonical variables are
derived. The Laplace transforms of these are determined in Sec.\
\ref{sectionlaplace}. As we shall see, it is essential to introduce forward
and backward Laplace transforms for positive and negative time,
respectively. Subsequently, in Sec.\ \ref{sectiongreen}, the Green
functions of the inhomogeneous dielectric are employed to determine a
space- and frequency-dependent source density, which is the analogue of the
noise-current density in the phenomenological quantization scheme. The
explicit form for this source density as a function of the canonical
variables is derived in Sec.\ \ref{sectionevaluation}. Once identified in
terms of the canonical variables, some important properties of the source
density can be derived, as presented in Sec. \ref{sectionproperties}. In
particular, we shall demonstrate that the damped-polariton model can be
diagonalized in terms of the source density. The full time dependence of
the electromagnetic field and of the dielectric polarization density can
thus be established. The paper ends with a discussion and with some
conclusions.\\

\section{Equations of motion\label{sectioneqmotion}}
As a model for an absorptive dielectric interacting with the
electromagnetic field we adopt the inhomogeneous damped-polariton
system. In this model the polarization density is a continuous
space-dependent variable. The damping is provided through the coupling to a
bath of harmonic oscillators with a continuous range of
eigenfrequencies. The bath coupling constant depends both on the frequency
and on the position. The electromagnetic field is coupled to the
polarization according to the standard minimal-coupling scheme.

The Lagrangian density of the damped-polariton system is \cite{HB92b}:
\begin{eqnarray}
\mathcal{L}&=&
\half \varepsilon_0 E^2-\half \mu_0^{-1} B^2 
+\half \rho \dot{X}^2 -\half \rho \omega_0^2 X^2 \nonumber\\
&&+\half\rho\int_0^{\infty}{\rm d}\omega\,\dot{Y}_{\omega}^2
-\half\rho\int_0^{\infty}{\rm d}\omega\, \omega^2\, Y_{\omega}^2\nonumber\\
&&-\Phi{\bm \nabla}\cdot(\alpha {\bf X})
-\alpha {\bf A}\cdot\dot{\bf X}
-\int_0^{\infty} {\rm d}\omega \, v_\omega\, {\bf X}\cdot\dot{\bf Y}_{\omega}.
\label{2.1}
\end{eqnarray}
The electromagnetic field is described by the scalar potential $\Phi({\bf
r})$ and the vector potential ${\bf A}({\bf r})$, with ${\bf
E}=-{\bm\nabla}\Phi-\dot{\bf A}$ and ${\bf B}=\nabla\times{\bf A}$. We
choose the Coulomb gauge in which the vector potential is purely
transverse, so that $\nabla\cdot{\bf A}=0$.  The dielectric degrees of
freedom are described by a space-dependent harmonic variable ${\bf X}({\bf
r})$, with an associated eigenfrequency $\omega_0({\bf r})$ and a density
$\rho({\bf r})$. The polarization density is given by $-\alpha {\bf X}$,
with $\alpha({\bf r})$ a space-dependent proportionality constant. In the
Coulomb gauge the scalar potential is a material variable that is given as
the solution of the Poisson equation $\Delta\Phi=
-\varepsilon_0^{-1}\nabla\cdot(\alpha{\bf X})$, with suitable boundary
conditions at infinity. The electromagnetic potentials $\Phi$, ${\bf A}$
and the dielectric variable ${\bf X}$ are coupled in the usual way, with
${\bm \nabla}\cdot(\alpha {\bf X})$ the bound charge density and
$-\alpha\dot{\bf X}$ the bound current density. Hence, $\alpha({\bf r})$
gives the strength of the coupling between the electromagnetic fields (or
the potentials) and ${\bf X}({\bf r})$. Damping is introduced in the model
through a continuum of harmonic-oscillator bath variables ${\bf
Y}_{\omega}({\bf r})$, labeled by the frequency $\omega$. The coupling to
${\bf X}({\bf r})$ is determined by the bath coupling parameter
$v_{\omega}({\bf r})$. A schematic representation of the system and its
independent parameters is:
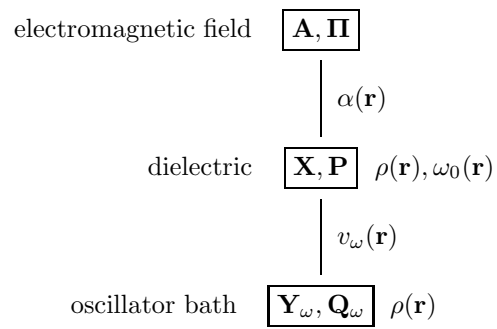
\begin{eqnarray}
\text{electromagnetic field}&
\fbox{${\displaystyle{{\bf A}, {\bm \Pi}}}$}&\nonumber\\
&\rule{0.2mm}{1cm}& \rule{-7mm}{0mm}\raisebox{4mm}{$\alpha({\bf r})$}
\nonumber\\
\text{dielectric}&
\fbox{${\displaystyle{{\bf X}, {\bf P}}}$}&
\rule{-2mm}{0mm}\rho({\bf r}),\omega_0({\bf r})
\nonumber\\
&\rule{0.2mm}{1cm}&\rule{-7mm}{0mm}\raisebox{4mm}{$v_{\omega}({\bf r})$}
\nonumber\\
\text{oscillator bath}\rule{2mm}{0mm}&
\fbox{${\displaystyle{{\bf Y}_{\omega}, {\bf Q}_{\omega}}}$}&
\rho({\bf r})
\nonumber
\end{eqnarray}
Here ${\bm \Pi}$, ${\bf P}$ and ${\bf Q}_{\omega}$ are canonical momenta,
which will be defined below.

In writing (\ref{2.1}) we have used the same conventions as in
\cite{HB92b}. In particular, we have refrained from a rescaling of the
physical variables. As is clear from (\ref{2.1}), the density $\rho$ could
have been scaled away by redefining ${\bf X}$ and ${\bf Y}_{\omega}$. In
this way we would have been left with the independent (rescaled) coupling
parameters $\alpha$ and $v_{\omega}$ (and the frequency
$\omega_0$). Turning the argument the other way around, we could as well
have chosen different density parameters $\rho_X$ and $\rho_Y$ multiplying
the contributions with ${\bf X}$ and ${\bf Y}_{\omega}$. In that way a
model would have been introduced that seems more general, but it is not.

Introducing the canonical momenta 
\begin{subequations}
\label{2.2}
\begin{eqnarray}
{\bm \Pi}&=&\frac{\partial \mathcal{L}}{\partial \dot{\bf A}}=\varepsilon_0 \, 
\dot{\bf A},\label{2.2a} \\
{\bf P}&=&\frac{\partial \mathcal{L}}{\partial \dot{\bf X}}
=\rho\,\dot{\bf X}-\alpha {\bf A},\label{2.2b} \\
{\bf Q}_{\omega}&=&\frac{\delta \mathcal{L}}{\delta \dot{\bf Y}_{\omega}}
=\rho {\dot{\bf Y}}_{\omega}-v_\omega\, {\bf X},\label{2.2c}
\end{eqnarray}
\end{subequations}
with $\delta$ a functional derivative in the variable $\omega$, we find the
Hamiltonian as
\begin{eqnarray}
H&=&\int {\rm d}{\bf r}\left[\frac{1}{2\epsilon_0} \Pi^2+
\frac{1}{2 \mu_0} ({\bm \nabla}\times {\bf A})^2 \right. \nonumber\\
&&+\frac{1}{2\rho}\, P^2+\half \rho\tilde{\omega}_0^2\, X^2\nonumber\\
&&+\frac{1}{2\rho}\int_0^{\infty}{\rm d}\omega\, Q_{\omega}^2 +
\half\rho\int_0^{\infty}{\rm d}\omega\, \omega^2\, Y_{\omega}^2\nonumber\\
&&\left. +\frac{\alpha}{\rho}{\bf A}\cdot{\bf P}+\frac{\alpha^2}{2\rho}\, A^2
+\frac{1}{\rho}\int_0^{\infty}{\rm d}\omega\, v_{\omega}\, {\bf X}\cdot{\bf
Q}_{\omega}
\right]\nonumber\\
&&+\int{\rm d}{\bf r}{\rm d}{\bf r}'\,
\frac{{\bm \nabla}\cdot(\alpha{\bf X})\, {\bm \nabla'}\cdot(\alpha'{\bf X}')}
{8\pi\epsilon_0|{\bf r}-{\bf r}'|}.\label{2.3}
\end{eqnarray}
We introduced the notation ${\bf X}'={\bf X}({\bf r}')$, and likewise $\alpha'$
and ${\bm \nabla'}$. The renormalized frequency $\tilde{\omega}_0({\bf r})$ is
defined by $\tilde{\omega}_0^2=\omega_0^2+\rho^{-2} \int_0^{\infty}{\rm
d}\omega\,v_\omega^2$, where the integral is assumed to be finite at all
positions ${\bf r}$.

To quantize the model we impose the usual commutation relations
\begin{subequations}
\label{2.4}
\begin{eqnarray}
\left[{\bm \Pi}({\bf r}),{\bf A}({\bf r}')\right]& = & 
-i\,\hbarit\, {\bm \delta}_{\rm T}({\bf r}-{\bf r}'),\label{2.4a}\\ 
\left[{\bf P}({\bf r}),{\bf X}({\bf r}')\right]& = & 
-i\,\hbarit\, \bfsfI \, \delta({\bf r}-{\bf r}'),\label{2.4b}\\ 
\left[{\bf Q}_{\omega}({\bf r}),{\bf Y}_{\omega'}({\bf r}')\right]& = &
-i\,\hbarit\, \delta(\omega-\omega')\, \bfsfI \,\delta({\bf r}-{\bf r}'),
\label{2.4c}
\end{eqnarray}
\end{subequations}
while all other commutators of the canonical variables vanish. Here $\bfsfI$ is
the three-dimensional unit tensor, while ${\bm
\delta}_{\rm T}({\bf r})=\bfsfI\, \delta({\bf r})+{\bm\nabla}{\bm\nabla}(4\pi
r)^{-1}$ is the transverse delta function.

With these `primary' canonical commutation relations and the relations between
the potentials and field operators, one can derive the following `secondary'
commutation relations
\begin{subequations}\label{2.5}
\begin{eqnarray}
\left[ {\bf E}({\bf r}),{\bf A}({\bf r'})\right] & = & \frac{i
\hbarit}{\varepsilon_{0}}\, {\bm \delta}_{\rm T}({\bf r}-{\bf r'}), \label{2.5a}\\
\left[ E_{j}({\bf r}),B_{k}({\bf r'})\right] & = & \frac{i
\hbarit}{\varepsilon_{0}}\,\epsilon_{jkl}\,\nabla'_l\delta({\bf
r}-{\bf r'}), \label{2.5b}
\end{eqnarray}
\end{subequations}
where we used (\ref{2.2a}). Notice that these commutation relations are
medium-independent, because in Eq.~(\ref{2.4}) we took the electromagnetic
field to be canonically independent from the material variables ${\bf X}$
and ${\bf P}$.

In the Heisenberg picture the equations of motion for the canonical variables
follow by evaluating the commutators with the Hamiltonian:
\begin{subequations}
\label{2.6}
\begin{eqnarray}
\dot{\bf A}&=&\frac{1}{\varepsilon_0}\, {\bm \Pi},\label{2.6a}\\
\dot{\bm \Pi}&=&\frac{1}{\mu_0}\, \Delta{\bf A}
-\left[\frac{\alpha}{\rho}\, ({\bf P}+\alpha{\bf A})\right]_{\rm T}, 
\label{2.6b}\\
\dot{\bf X}&=&\frac{1}{\rho}\, ({\bf P}+\alpha{\bf A}),\label{2.6c}\\
\dot{\bf P}&=&-\rho \, \tilde{\omega}_0^2\, {\bf X}
-\frac{\alpha}{\varepsilon_0}\,[\alpha {\bf X}]_{\rm L}
-\frac{1}{\rho}\int_0^{\infty}{\rm d}\omega\, v_{\omega} \, {\bf Q}_{\omega},
\label{2.6d}\\
\dot{\bf Y}_{\omega}&=&\frac{1}{\rho}\,({\bf Q}_{\omega}+v_{\omega}\, {\bf
X}),\label{2.6e}\\
\dot{\bf Q}_{\omega}&=&-\rho\, \omega^2\, {\bf Y}_{\omega}.\label{2.6f}
\end{eqnarray}
\end{subequations}
In the second equation the subscript ${\rm T}$ denotes the transverse part
of the vector, which is obtained by a convolution with the transverse delta
function. Likewise, the subscript ${\rm L}$ in the fourth equation
indicates the longitudinal part, obtained by convolving with the
longitudinal delta function ${\bm \delta}_{\rm L}({\bf
r})=-{\bm\nabla}{\bm\nabla}(4\pi r)^{-1}$.

From (\ref{2.6a})-(\ref{2.6c}) one gets
\begin{equation}
\Delta{\bf A}-c^{-2}\,\ddot{\bf A}=\mu_0\, \left[\frac{\alpha}{\rho}\, 
({\bf P}+\alpha {\bf A})\right]_{\rm T}=\mu_0\, [\alpha\dot{\bf X}]_{\rm T}.
\label{2.7}
\end{equation}
This equation is equivalent to Maxwell's equation
\begin{equation}
-{\bm \nabla}\times{\bf B}+c^{-2}\,\dot{\bf E}=\mu_0\, \alpha \dot{\bf X},
\label{2.8}
\end{equation}
since the Poisson equation for $\Phi$ can be rewritten as 
\begin{equation}
[\alpha {\bf X}]_{\rm L}=-\varepsilon_0 {\bm\nabla}\Phi. \label{2.9}
\end{equation}

The equations of motion (\ref{2.6c})-(\ref{2.6d}) for the variables of the
dielectric medium yield a second-order differential equation for ${\bf X}$:
\begin{equation}
\rho\ddot{\bf X}+\rho\tilde{\omega}_0^2{\bf X}=\alpha\dot{\bf A}
-\frac{\alpha}{\varepsilon_0}\, [\alpha{\bf X}]_{\rm L}
-\frac{1}{\rho}\int_0^{\infty}{\rm d}\omega\, v_{\omega}\, {\bf
Q}_{\omega}, \label{2.10}
\end{equation}
or alternatively, with the use of (\ref{2.6e}):
\begin{equation}
\rho\ddot{\bf X}+\rho{\omega}_0^2{\bf X}
=\alpha\dot{\bf A}
-\frac{\alpha}{\varepsilon_0}\, [\alpha{\bf X}]_{\rm L}
-\int_0^{\infty}{\rm d}\omega\, v_{\omega}\, \dot{\bf Y}_{\omega}.
\label{2.11}
\end{equation}
The change from $\tilde{\omega}_0$ to $\omega_0$ should be noted here. The
first two terms at the right-hand sides of (\ref{2.10}) and (\ref{2.11}) 
are equal to $-\alpha{\bf E}$, as follows from (\ref{2.9}).

Finally, the second-order equation of motion for the bath follows from
(\ref{2.6e})-(\ref{2.6f}) as
\begin{equation}
\rho\ddot{\bf Y}_{\omega}+\rho\,\omega^2{\bf Y}_{\omega}=
v_{\omega}\, \dot{\bf X}. \label{2.12}
\end{equation}
The second-order equations (\ref{2.7}), (\ref{2.11}) and (\ref{2.12})
determine the time evolution of the basic physical variables ${\bf A}$,
${\bf X}$ and ${\bf Y}_{\omega}$, which represent the vector potential, the
dielectric polarization density and the harmonic-oscillator displacement
density of the bath. If the initial conditions of ${\bf A}$, ${\bf X}$ and
${\bf Y}_{\omega}$ and their first time derivatives are given at $t=0$,
these operators are known at any value of $t$, either positive or
negative.\\

\section{Laplace transforms\label{sectionlaplace}}

Our goal for the rest of this paper is twofold. First, we would like to
determine the dynamics of the model and to find the complete time
dependence of the canonical variables. Secondly, we want to show how the
phenomenological theory emerges from our model. To that end, we must
identify the elements of the phenomenological theory (dielectric function,
Green function, and noise-current density) in terms of the
variables from the model. Then, after their identification we must show
that these quantities have all the desired properties that are merely
postulated in the phenomenological approach.

The equations of motion of the previous section constitute a set of linear
differential equations. In this section, we solve them in terms of their
initial conditions, by introducing Laplace transforms. This technique has
been used before in the analysis of the damped-polariton model
\cite{WS01,JRH03}. In particular, we will derive an equation for the
Laplace transform of the electric field ${\bf E}$, while taking care to
eliminate the Laplace-transformed variables describing the dielectric
medium. To that end we shall first solve the Laplace-transformed equations
for the bath variables.  Subsequently, these will be used to obtain an
identity that relates the Laplace transforms of the polarization density
and the electric field. While establishing that relationship, we shall
identify the electric susceptibility in Laplace language. Finally, wave
equations for the Laplace transforms of the electric field will be
deduced. All equations will be valid for an arbitrary spatial dependence of
the variables of the model.

For any time-dependent operator $\Omega$ the (forward) Laplace transform is
defined as
\begin{equation}
\bar{\Omega}(p)=\int_0^{\infty}{\rm d}t\, e^{-pt}\, \Omega(t).\label{3.1}
\end{equation}
Obviously, the Laplace transform contains all information on the time
evolution of $\Omega$ for positive $t$. In the following we wish to
determine the time evolution of the relevant operators of our model for any
time, either positive or negative. Hence, we also introduce the backward
Laplace transform:
\begin{equation}
\breve{\Omega}(p)=\int_0^{\infty}{\rm d}t\, e^{-pt}\, \Omega(-t).\label{3.2}
\end{equation}
Both transforms are defined for all $p$ with ${\rm Re}\, p>0$. 

Carrying out the (forward) Laplace transformation of (\ref{2.6e}) and
(\ref{2.6f}), and eliminating $\bar{\bf Y}_{\omega}(p)$ we find
\begin{eqnarray}
\bar{\bf Q}_{\omega}(p)&=&-\frac{\omega^2}{p^2+\omega^2}\, 
v_{\omega}\,\bar{\bf X}(p)\nonumber\\
&&+\frac{1}{p^2+\omega^2}\left[p\,{\bf Q}_{\omega}(0)
-\rho\,\omega^2\,{\bf Y}_{\omega}(0)\right], \label{3.3}
\end{eqnarray}
with ${\bf Q}_{\omega}(0)$ and ${\bf Y}_{\omega}(0)$ the initial conditions
at $t=0$. Alternatively, we could have used (\ref{2.12}) as a starting
point. Upon performing its Laplace transformation, we may eliminate the initial
condition for $\dot{\bf Y}_{\omega}$ with the help of (\ref{2.6e}). The
initial condition of ${\bf X}$ is then found to drop out as well, so that
(\ref{3.3}) is recovered.

From (\ref{2.6c})-(\ref{2.6d}) we find analogously, after elimination of
$\bar{\bf P}(p)$:
\begin{eqnarray}
\left(p^2+\tilde{\omega}_0^2\right)\, \bar{\bf X}(p)&=&
\frac{\alpha}{\rho}\left\{ p\bar{\bf A}(p)-
\frac{1}{\varepsilon_0}[\alpha\bar{\bf
X}(p)]_{\rm L}\right\}\nonumber\\
&&\rule{-3cm}{0cm}
-\frac{1}{\rho^2}
\int_0^{\infty}{\rm d}\omega\, v_{\omega}\bar{\bf Q}_{\omega}(p)
+p{\bf X}(0)+\frac{1}{\rho}{\bf P}(0). \label{3.4}
\end{eqnarray}
Insertion of (\ref{3.3}) in the integral yields:
\begin{eqnarray}
\left[p^2+\tilde{\omega}_0^2
-\frac{1}{\rho^2}\int_0^{\infty}{\rm d}\omega\, 
\frac{\omega^2\, v_{\omega}^2}{p^2+\omega^2}\right]\, \bar{\bf X}(p)
&&\nonumber\\
&&\rule{-6cm}{0cm}= 
\frac{\alpha}{\rho}\left\{ p\bar{\bf A}(p)-\frac{1}{\varepsilon_0}
[\alpha\bar{\bf X}(p)]_{\rm L}\right\} +p{\bf X}(0)+\frac{1}{\rho}{\bf P}(0)
\nonumber\\
&&\rule{-6cm}{0cm}+\frac{1}{\rho}\int_0^{\infty}{\rm d}\omega\, 
\frac{v_{\omega}}{p^2+\omega^2}\, 
\left[\omega^2\, {\bf Y}_{\omega}(0)
-\frac{p}{\rho}\,{\bf Q}_{\omega}(0)\right]. \label{3.5}
\end{eqnarray}
The first few terms at the right-hand side are related to the Laplace transform
of the electric field, which reads
\begin{equation}
\bar{\bf E}(p)=-p\bar{\bf A}(p)+\frac{1}{\varepsilon_0}[\alpha\bar{\bf
X}(p)]_{\rm L}+{\bf A}(0). \label{3.6}
\end{equation}

From (\ref{3.5}) with (\ref{3.6}) it follows that the Laplace-transformed
polarization density $-\alpha\bar{\bf X}(p)$ is proportional to the
Laplace-transformed electric field, apart from terms depending on the
initial conditions. We identify the proportionality constant as the
electric susceptibility in Laplace language:
\begin{equation}
\bar{\chi}(p)=\frac{\alpha^2}{\varepsilon_0\rho}\, 
\frac{1}{p^2+\tilde{\omega}_0^2
-\rho^{-2}\int_0^{\infty}{\rm d}\omega\, 
\omega^2\, v_{\omega}^2/(p^2+\omega^2)} . \label{3.7}
\end{equation}
All parameters at the right-hand side depend on position, so that the
susceptibility is a space-dependent quantity. In this respect it is a
generalization of the definitions in \cite{HB92a,HB92b,DF98a,DF98b}, which
are valid for homogeneous dielectrics. The susceptibility is an analytic
function of $p$ for all $p$ with ${\rm Re}\, p>0$. Indeed, one can prove
that the denominator in (\ref{3.7}) cannot vanish for any $p$ in the right
half-plane. As a consequence of its analyticity properties, the
susceptibility satisfies the standard Kramers-Kronig relations which
connect the real and imaginary parts of $\bar{\chi}(p)$.

After introduction of the susceptibility $\bar{\chi}(p)$ the linear
relationship between $\bar{\bf X}(p)$ and $\bar{\bf E}(p)$ gets the form:
\begin{eqnarray}
\bar{\bf X}(p)&=&-\frac{\varepsilon_0}{\alpha}\, \bar{\chi}(p)\, \bar{\bf E}(p)
\nonumber\\
&&\rule{-1cm}{0cm}+\frac{\varepsilon_0}{\alpha^2}\, \bar{\chi}(p)\, 
\biggl\{\alpha\, {\bf A}(0)
+\rho p\, {\bf X}(0)+{\bf P}(0)\nonumber\\
&&\rule{-1cm}{0cm} +\int_0^{\infty}{\rm d}\omega\, 
\frac{v_{\omega}}{p^2+\omega^2}\, 
\left[\omega^2\, {\bf Y}_{\omega}(0)
-\frac{p}{\rho}\,{\bf Q}_{\omega}(0)\right]\biggr\}. \label{3.8}
\end{eqnarray}
This identity, which relates the Laplace transforms of the polarization
density and the electric field, will be crucial in eliminating the
dielectric variables, as we shall see presently. As expected, the
relationship still depends on the initial values of all canonical variables
(with the exception of ${\bm \Pi}(0)$).

Having succeeded in expressing the polarization density in terms of the
electric field, we want to establish a wave equation for the electric field
in Laplace language. All Laplace-transformed equations, which we derived
above, contain terms depending on the initial conditions. Hence, we expect
that the wave equation will contain such terms as well. In fact, we shall
show that owing to the presence of these terms, the wave equation will be
inhomogeneous.

To derive the wave equation in Laplace language, we have to use the
remaining field-dependent equations in the set (\ref{2.6}). However,
instead of (\ref{2.6a})-(\ref{2.6b}) we prefer to employ their corollary
(\ref{2.8}), or, better still, the equation which follows by taking the
time derivative of (\ref{2.8}) and using the induction law:
\begin{equation} 
{\bm \nabla}\times ({\bm \nabla}\times {\bf E})+c^{-2}\ddot{\bf E}=
\mu_0\alpha\, \ddot{\bf X}. \label{3.9}
\end{equation}
After a (forward) Laplace transformation we get
\begin{eqnarray}
{\bm \nabla}\times [{\bm \nabla}\times \bar{\bf E}(p)]
+c^{-2}p^2\bar{\bf E}(p)-\mu_0 \alpha p^2\, \bar{\bf X}(p)&&\nonumber\\
&&\rule{-7.5cm}{0cm}= c^{-2}\, \dot{\bf E}(0)+ c^{-2}p\, {\bf E}(0)
-\mu_0 \alpha \,\dot{\bf X}(0)-\mu_0\alpha  p \,{\bf X}(0). 
\label{3.10}
\end{eqnarray}
As before, the introduction of Laplace transforms has led to terms
depending on the initial conditions. We wish to express these in terms of
the canonical variables at $t=0$. At the right-hand side we use
(\ref{2.6c}) for $t=0$ to rewrite $\dot{\bf X}(0)$:
\begin{equation}
\dot{\bf X}(0)=\frac{\alpha}{\rho}\,{\bf A}(0)
+\frac{1}{\rho}\,{\bf P}(0) .\label{3.11}
\end{equation}
Upon inserting this relation in (\ref{2.8}) for $t=0$ we find
\begin{equation}
\dot{\bf E}(0)=c^2\,{\bm \nabla}\times [{\bm \nabla}\times{\bf A}(0)]
+\frac{\alpha^2}{\varepsilon_0\rho}\,{\bf A}(0)
+\frac{\alpha}{\varepsilon_0\rho}\,{\bf P}(0).\label{3.12}
\end{equation}
Furthermore, we use (\ref{2.6a}) and (\ref{2.9}) to write the electric
field at $t=0$ as
\begin{equation}
{\bf E}(0)=-\frac{1}{\varepsilon_0}{\bm \Pi}(0)+
\frac{1}{\varepsilon_0}[\alpha{\bf X}(0)]_{\rm L}. \label{3.13}
\end{equation}
Substituting (\ref{3.11})-(\ref{3.13}) in the right-hand side of
(\ref{3.10}), and using the expression (\ref{3.8}) for $\bar{\bf X}(p)$ in
the last term at the left-hand side, we finally arrive at the
Laplace-transformed wave equation for the electric field in the form:
\begin{equation}
{\bm \nabla}\times [{\bm \nabla}\times \bar{\bf E}(p)]
+c^{-2}p^2\bar{\varepsilon}(p)\bar{\bf E}(p)=-\mu_0 p\, \bar{\bf J}(p) , 
\label{3.14}
\end{equation} 
with $\bar{\epsilon}(p)=1+\bar{\chi}(p)$ the (relative) electric
permeability in Laplace language. 

The differential equation (\ref{3.14}) in Laplace language is an
inhomogeneous wave equation. The source term at the right-hand side depends
on the initial conditions at time $t=0$:
\begin{eqnarray}
\bar{\bf J}(p)&=&
-\frac{1}{\mu_0 p}\, {\bm \nabla}\times [{\bm \nabla}\times{\bf A}(0)]
-\varepsilon_0 p \bar{\chi}(p)\, {\bf A}(0)+{\bm \Pi}(0)\nonumber\\
&&\rule{-1.5cm}{0cm}+\alpha\left[1-\frac{\varepsilon_0\rho}{\alpha^2} 
p^2\bar{\chi}(p)\right]\, {\bf X}(0)
-[\alpha {\bf X}(0)]_{\rm L}
-\frac{\varepsilon_0}{\alpha}p\bar{\chi}(p)\, {\bf  P}(0)\nonumber\\
&&\rule{-1.5cm}{0cm}-\frac{\varepsilon_0}{\alpha} p \bar{\chi}(p)
\int_0^{\infty}{\rm d}\omega\, 
\frac{v_{\omega}}{p^2+\omega^2}\, 
\left[\omega^2\, {\bf Y}_{\omega}(0)
-\frac{p}{\rho}\,{\bf Q}_{\omega}(0)\right]. \label{3.15}
\end{eqnarray}

In principle, the electric field in the inhomogeneous wave equation
(\ref{3.14}) can be solved in terms of its source. Upon performing the
inverse Laplace transformation, one then gets an expression for the
electric field at any later time after $t=0$. However, as noted above, we
also need the electric field (and the other variables of the model) for all
times prior to $t=0$. To find that information we have to derive the
corresponding wave equation for the backward Laplace transform of the
electric field. We shall proceed along the same lines as above.

The backward Laplace transforms of (\ref{2.6e}) and (\ref{2.6f}) yield on a
par with (\ref{3.3}):
\begin{eqnarray}
\breve{\bf Q}_{\omega}(p)&=&-\frac{\omega^2}{p^2+\omega^2}\, 
v_{\omega}\,\breve{\bf X}(p)\nonumber\\
&&+\frac{1}{p^2+\omega^2}\left[p\,{\bf Q}_{\omega}(0)
+\rho\,\omega^2\,{\bf Y}_{\omega}(0)\right], \label{3.16}
\end{eqnarray}
where the change of sign in the last term should be noted.  Using this
equation and following the same steps as above, we find the counterpart of
(\ref{3.8}) as
\begin{eqnarray}
\breve{\bf X}(p)&=&-\frac{\varepsilon_0}{\alpha}\, \bar{\chi}(p)\, 
\breve{\bf E}(p)
\nonumber\\
&&\rule{-1cm}{0cm}+\frac{\varepsilon_0}{\alpha^2}\, \bar{\chi}(p)\, 
\biggl\{-\alpha\, {\bf A}(0)
+\rho p\, {\bf X}(0)-{\bf P}(0)\nonumber\\
&&\rule{-1cm}{0cm} -\int_0^{\infty}{\rm d}\omega\, 
\frac{v_{\omega}}{p^2+\omega^2}\, 
\left[\omega^2\, {\bf Y}_{\omega}(0)
+\frac{p}{\rho}\,{\bf Q}_{\omega}(0)\right]\biggr\}. \label{3.17}
\end{eqnarray}
The susceptibility is given by (\ref{3.7}), as before. The backward Laplace
transform of the electric field is slightly different from (\ref{3.6}):
\begin{equation}
\breve{\bf E}(p)=p\breve{\bf A}(p)+\frac{1}{\varepsilon_0}[\alpha\breve{\bf
X}(p)]_{\rm L}-{\bf A}(0). \label{3.18}
\end{equation}
Here, as in (\ref{3.17}), several terms have changed sign. Finally, turning
to the field equation, we obtain from (\ref{3.9}) after a backward Laplace
transformation:
\begin{eqnarray}
{\bm \nabla}\times [{\bm \nabla}\times \breve{\bf E}(p)]
+c^{-2}p^2\breve{\bf E}(p)-\mu_0 \alpha p^2\, \breve{\bf X}(p)&&\nonumber\\
&&\rule{-7.5cm}{0cm}= -c^{-2}\, \dot{\bf E}(0)+ c^{-2}p\, {\bf E}(0)
+\mu_0 \alpha \,\dot{\bf X}(0)-\mu_0\alpha  p \,{\bf X}(0). 
\label{3.19}
\end{eqnarray}
Substitution of (\ref{3.11})-(\ref{3.13}) yields as the analogue of
(\ref{3.14}):
\begin{equation}
{\bm \nabla}\times [{\bm \nabla}\times \breve{\bf E}(p)]
+c^{-2}p^2\bar{\varepsilon}(p)\breve{\bf E}(p)=\mu_0 p\, \breve{\bf J}(p) .
\label{3.20}
\end{equation} 
The source term in this inhomogeneous wave equation is:
\begin{eqnarray}
\breve{\bf J}(p)&=&
-\frac{1}{\mu_0 p}\, {\bm \nabla}\times [{\bm \nabla}\times{\bf A}(0)]
-\varepsilon_0 p \bar{\chi}(p)\, {\bf A}(0)-{\bm \Pi}(0)\nonumber\\
&&\rule{-1.5cm}{0cm}-\alpha\left[1-\frac{\varepsilon_0\rho}{\alpha^2} 
p^2\bar{\chi}(p)\right]\, {\bf X}(0)
+[\alpha {\bf X}(0)]_{\rm L}
-\frac{\varepsilon_0}{\alpha}p\bar{\chi}(p)\, {\bf  P}(0)\nonumber\\
&&\rule{-1.5cm}{0cm}-\frac{\varepsilon_0}{\alpha} p \bar{\chi}(p)
\int_0^{\infty}{\rm d}\omega\, 
\frac{v_{\omega}}{p^2+\omega^2}\, 
\left[\omega^2\, {\bf Y}_{\omega}(0)
+\frac{p}{\rho}\,{\bf Q}_{\omega}(0)\right]. \label{3.21}
\end{eqnarray}
As expected, several terms have changed sign as compared to
(\ref{3.15}). For future convenience we have chosen the sign in the
right-hand side of (\ref{3.20}) to be the opposite of that in (\ref{3.14}).

The main results in this section are the expressions (\ref{3.15}) and
(\ref{3.21}) for $\bar{\bf J}(p)$ and $\breve{\bf J}(p)$. These are the
source terms of the Laplace-transformed wave equations (\ref{3.14}) and
(\ref{3.20}). It should be noted that these source terms are {\em not}
defined as the forward and backward Laplace transforms of some operator
${\bf J}({\bf r},t)$, although their notation might suggest otherwise. As a
consequence, their properties differ from other pairs of operators, like
$\bar{\bf E}(p)$ and $\breve{\bf E}(p)$. In Sec.\ \ref{sectionproperties}
we shall come back to this point.\\

\section{Green functions and solutions of wave equations\label{sectiongreen}}
In the previous section we have seen that both the forward and the backward
Laplace transform of the electric field satisfy a wave equation with a
source term. To solve these equations we introduce tensorial Green
functions in Laplace language. The Green function associated to the wave
equations (\ref{3.14}) and (\ref{3.20}) is defined as the solution of the
differential equation \cite{DKW98,SKW98,KSW01}:
\begin{eqnarray}
-{\bm \nabla}\times [{\bm \nabla}\times \bar{\bfsfG}({\bf r},{\bf r}',p)]
-\frac{p^2}{c^2}\bar{\varepsilon}({\bf r},p)\, \bar{\bfsfG}({\bf r},{\bf r}',p)
\nonumber\\
=\bfsfI\, \delta({\bf r}-{\bf r}'), \label{4.1}
\end{eqnarray} 
here for clarity we reintroduced the spatial argument of the
permeability. The function $\bar{\bfsfG}({\bf r},{\bf r}',p)$ is the
forward Laplace transform of the standard retarded Green function of
macroscopic electrodynamics. It is also equal to the backward Laplace
transform of the advanced Green function. The Green function is analytic
for all $p$ in the half-plane with ${\rm Re}\, p>0$, as is the
susceptibility $\bar{\chi}({\bf r},p)$ \cite{SKW98}. The defining equation
(\ref{4.1}) can be read as the statement that the Green function is the
inverse of the operator $ -[{\bm \nabla}\times ({\bm
\nabla}\times)]-c^{-2}p^2\bar{\varepsilon}({\bf r},p)(\bfsfI\cdot)$, which
is a symmetric differential operator in the space of square-integrable
vector functions. Owing to the symmetry the Green function satisfies the
reciprocity relation:
\begin{equation}
\left[\bar{\bfsfG}({\bf r},{\bf r}',p)\right]_{ij}=
\left[\bar{\bfsfG}({\bf r}',{\bf r},p)\right]_{ji}. \label{4.2}
\end{equation}
The adjoint equation of (\ref{4.1}) reads
\begin{eqnarray}
-\biggl[\bar{\bfsfG}({\bf r},{\bf r}',p)\times 
\overleftarrow{\bm  \nabla'}\biggr]\times\overleftarrow{\bm \nabla'}
-\frac{p^2}{c^2}\bar{\varepsilon}({\bf r}',p)\, 
\bar{\bfsfG}({\bf r},{\bf r}',p)
\nonumber\\
=\bfsfI\, \delta({\bf r}-{\bf r}'), 
\label{4.3}
\end{eqnarray} 
where the spatial derivatives operate to the left. 

In terms of the Green function, the solution of (\ref{3.14}) for the Laplace
transform of the electric field is
\begin{equation}
\bar{\bf E}({\bf r},p)=\mu_0 p \int d{\bf r}'\, 
\bar{\bfsfG}({\bf r},{\bf r}',p)\cdot
\bar{\bf J}({\bf r}',p). \label{4.4}
\end{equation}
Likewise, the backward Laplace transform of the field follows from
(\ref{3.20}) as
\begin{equation}
\breve{\bf E}({\bf r},p)=-\mu_0 p \int d{\bf r}'\, 
\bar{\bfsfG}({\bf r},{\bf r}',p)\cdot \breve{\bf J}({\bf r}',p). \label{4.5}
\end{equation}
The two integral representations (\ref{4.4}) and (\ref{4.5}) for the
forward and the backward Laplace transform of the electric field contain
all information that is needed to express the electric field at time $t$ in
terms of the initial conditions of the canonical variables. The latter show
up explicitly when the expressions (\ref{3.15}) and (\ref{3.21}) are
substituted in the integrals.

The time-dependent electric field is obtained from the integral
representations by an inverse Laplace transformation. From (\ref{4.4}) we
get the electric field for $t>0$:
\begin{eqnarray}
{\bf E}({\bf r},t)&=&-\frac{i\mu_0}{2\pi} \int_{-\infty}^{\infty} 
d\omega\, e^{-i\omega t}\, 
\omega \nonumber\\
&&\rule{-1cm}{0cm}
\times\int d{\bf r}'\, \bar{\bfsfG}({\bf r},{\bf r}',-i\omega +0)\cdot
\bar{\bf J}({\bf r}',-i\omega+0). \label{4.6}
\end{eqnarray}
Here we changed the integration variable from $p$ in the right half-plane
to $-i\omega +\eta$, with a small but positive $\eta$. We formally replace
$\eta$ by $0$, so that $-i\omega +\eta$ becomes $-i\omega +0$.

The electric field for $t<0$ is obtained from the inverse Laplace transform
of (\ref{4.5}):
\begin{eqnarray}
{\bf E}({\bf r},t)&=&-\frac{i\mu_0}{2\pi} \int_{-\infty}^{\infty} 
d\omega\, e^{-i\omega t}\, 
\omega \nonumber\\
&&\times\int d{\bf r}'\, \bar{\bfsfG}({\bf r},{\bf r}',i\omega +0)\cdot
\breve{\bf J}({\bf r}',i\omega+0). \label{4.7}
\end{eqnarray}

The $\omega$-dependent integrand in (\ref{4.6}) is analytic for all
$\omega$ in the upper half-plane. Hence, the integral over $\omega$
vanishes for negative $t$, since the $\omega$-contour can be closed in the
upper half-plane for $t<0$. Likewise, the right-hand side of (\ref{4.7}) is
zero for positive $t$. As a consequence, one may combine the two
expressions into a single one, which is valid for all $t$:
\begin{equation}
{\bf E}({\bf r},t)=\int_0^{\infty}d\omega\, e^{-i\omega t}\, 
{\bf E}^{(+)}({\bf r},\omega)+\text{h.c.}, \label{4.8}
\end{equation}
with the positive-frequency Fourier component:
\begin{eqnarray}
{\bf E}^{(+)}({\bf r},\omega)&=&-\frac{i\mu_0 \omega}{2\pi}
\nonumber\\ 
&&\rule{-2cm}{0cm}\times\int d{\bf r}'
\bigl[\bar{\bfsfG}({\bf r},{\bf r}',-i\omega +0)\cdot
\bar{\bf J}({\bf r}',-i\omega+0)\nonumber\\
&& +\bar{\bfsfG}({\bf r},{\bf r}',i\omega +0)\cdot
\breve{\bf J}({\bf r}',i\omega+0)\bigr].\label{4.9}
\end{eqnarray}
Alternatively, one may write
\begin{equation}
 {\bf E}^{(+)}({\bf r},\omega)=\frac{1}{2\pi}
\left[\bar{\bf E}({\bf r},-i\omega+0)
+\breve{\bf E}({\bf r},i\omega+0)\right], \label{4.10}
\end{equation}
as follows by going back to (\ref{4.4}) and (\ref{4.5}).  

We have succeeded now in obtaining the full time dependence of the
electric-field operator. As the representation (\ref{4.8}) shows, the field
is a linear superposition of contributions, each with its own time
dependence, and with a weight that is determined by the positive-frequency
Fourier component (\ref{4.9}). The latter is itself a linear combination
of the canonical variables at $t=0$, as follows from (\ref{3.15}) and
(\ref{3.21}).

The representation (\ref{4.8}) is valid for all $t$. Hence, consistency
demands that the right-hand side of (\ref{4.8}) should reduce to the
electric field (\ref{3.13}) in the limit $t\rightarrow 0$. In Appendix A we
show that this is indeed the case. It turns out that the proof depends on
the validity of several sum rules for the Green function. The latter hold
true as a consequence of the analyticity of the Green function and of its
asymptotic behavior for large frequencies.

Having determined the positive-frequency Fourier component of the
electric-field operator, we may try and find the differential equation
which it satisfies. Of course, one expects this equation to be of similar
form as in (\ref{3.14}) and (\ref{3.20}), with a frequency $\omega$ instead
of the Laplace variable $p$. Let us introduce therefore the permeability in
the frequency domain as $\varepsilon({\bf r},\omega)=\bar{\varepsilon}({\bf
r},-i\omega+0)$, for real $\omega$. Likewise, we will write
$\bar{\chi}({\bf r},-i\omega+0)$ as $\chi({\bf r},\omega)$. In terms of the
permeability $\varepsilon({\bf r},\omega)$ we may define the differential
operator $ -[{\bm \nabla}\times ({\bm
\nabla}\times)]+c^{-2}\omega^2\varepsilon({\bf
r},\omega)(\bfsfI\cdot)$. The positive-frequency Fourier component of the
electric field satisfies a differential equation containing this operator:
\begin{eqnarray}
-{\bm \nabla}\times\left[{\bm \nabla}\times 
{\bf E}^{(+)}({\bf r},\omega)\right] 
+\frac{\omega^2}{c^2}\varepsilon({\bf r},\omega)
{\bf E}^{(+)}({\bf r},\omega)&&\nonumber\\
&&\rule{-4cm}{0cm}= -i\mu_0\omega \, {\bf J}({\bf r},\omega). 
\label{4.11}
\end{eqnarray}
As expected, the differential equation is inhomogeneous, like (\ref{3.14})
and (\ref{3.20}). Apart from a trivial factor, the right-hand side contains
a source term ${\bf J}({\bf r},\omega)$. Since the positive-frequency part
${\bf E}^{(+)}({\bf r},\omega)$ of the electric field is known from
(\ref{4.9}) with (\ref{3.15}) and (\ref{3.21}), the differential equation
(\ref{4.11}) may serve as the definition of the operator ${\bf J}({\bf
r},\omega)$. Defining ${\bf J}({\bf r},\omega)$ in this way we try to make
contact with the phenomenological quantization procedure, in which an
equation of the same form as (\ref{4.11}) plays an important role
\cite{GW95, MLBJ95,GW96a,ML96,GW96b,DKW98,SKW98}. Although the equations
have the same form in the two theories, their interpretation is rather
different. In the phenomenological quantization procedure neither ${\bf
E}^{(+)}({\bf r},\omega)$ nor ${\bf J}({\bf r},\omega)$ are known at
first. To obtain a well-defined theory one has to postulate several
properties of ${\bf J}({\bf r},\omega)$. In particular, one postulates an
identity for the commutator of ${\bf J}({\bf r},\omega)$ with its hermitian
conjugate. Only after doing so does one arrive at a meaningful theory. In
the present theory we are able to {\em derive} the algebraic properties of
${\bf J}({\bf r},\omega)$, as we shall see later on.

As in the phenomenological theory, the operator ${\bf J}({\bf r},\omega)$
may be interpreted as a (frequency-dependent) noise-current density. Its
form still needs to be elaborated in more detail. By substituting
(\ref{4.9}) at the left-hand side of (\ref{4.11}) and using the definition
(\ref{4.1}) of the Green function to eliminate the differential operators,
we get
\begin{eqnarray}
{\bf J}({\bf r},\omega)&=&\frac{1}{2\pi}
\left[\bar{\bf J}({\bf r},-i\omega+0)
+\breve{\bf J}({\bf r},i\omega+0)\right]\nonumber\\
&&\rule{-1cm}{0cm}+\frac{\omega^2}{2\pi c^2}\left[\bar{\chi}({\bf r},
-i\omega+0)-
\bar{\chi}({\bf r},i\omega+0)\right]\nonumber\\
&&\rule{-1cm}{0cm}\times\int d{\bf r}'\, \bar{\bfsfG}({\bf r},{\bf r}',
i\omega +0)\cdot
\breve{\bf J}({\bf r}',i\omega+0).\label{4.12}
\end{eqnarray}
It turns out that the operator ${\bf J}({\bf r},\omega)$ has a more
complicated structure than the positive-frequency Fourier component
(\ref{4.10}) of the electric field. Whereas the first two terms at the
right-hand side have the expected form, an additional contribution shows
up, which has arisen from the second term of (\ref{4.9}). In fact, the
operator acting on the electric field in (\ref{4.11}) is the inverse of
$\bar{\bfsfG}({\bf r},{\bf r}',-i\omega+0)$, but {\em not} of
$\bar{\bfsfG}({\bf r},{\bf r}',i\omega+0)$: the frequency arguments of the
permeability do not match in the latter case. The additional term is
proportional to the imaginary part of the susceptibility $\bar{\chi}({\bf
r},-i\omega+0)$.

The operator ${\bf J}({\bf r},\omega)$ will play an important role in the
following. Before evaluating it explicitly in terms of the canonical
variables, we will show that it can also be obtained in a different
way. Let us consider, on a par with (\ref{4.10}), the positive-frequency
Fourier component of ${\bf X}$. As one might suppose that it is
proportional to ${\bf E}^{(+)}$, with a proportionality constant determined
by the susceptibility, we will focus on the combination
\begin{eqnarray}
-\alpha{\bf X}^{(+)}({\bf r},\omega)
-\varepsilon_0\chi({\bf r},\omega){\bf E}^{(+)}({\bf r},\omega)&&\nonumber\\
&&\rule{-6cm}{0cm}=-\frac{\alpha}{2\pi}
\left[\bar{\bf X}({\bf r},-i\omega+0)
+\breve{\bf X}({\bf r},i\omega+0)\right]\nonumber\\
&&\rule{-6cm}{0cm}-\frac{\varepsilon_0}{2\pi}\bar{\chi}({\bf r},-i\omega+0)
\left[\bar{\bf E}({\bf r},-i\omega+0)
+\breve{\bf E}({\bf r},i\omega+0)\right].\rule{0.8cm}{0cm}\label{4.13}
\end{eqnarray}
Comparing (\ref{3.10}) and (\ref{3.14}) and taking $p=-i\omega+0$ we infer
that the forward Laplace transforms satisfy the identity:
\begin{eqnarray}
-\alpha\bar{{\bf X}}(-i\omega+0)
-\varepsilon_0\bar{\chi}(-i\omega+0)\bar{\bf E}(-i\omega+0)
&&\nonumber\\
&&\rule{-6cm}{0cm}=\frac{i}{\omega}\bar{\bf J}(-i\omega+0)
-\frac{1}{\omega^2}\left[-\alpha\dot{\bf X}(0)
+\varepsilon_0\dot{\bf E}(0)\right]\nonumber\\
&&\rule{-6cm}{0cm}+\frac{i}{\omega}\left[-\alpha{\bf X}(0)
+\varepsilon_0{\bf E}(0)\right],\label{4.14}
\end{eqnarray}
where we suppressed the dependence on ${\bf r}$ for the moment. Likewise,
from (\ref{3.19})-(\ref{3.20}) we derive for the backward Laplace
transforms:
\begin{eqnarray}
-\alpha\breve{{\bf X}}(i\omega+0)
-\varepsilon_0\bar{\chi}(i\omega+0)\breve{\bf E}(i\omega+0)
&&\nonumber\\
&&\rule{-6cm}{0cm}=\frac{i}{\omega}\breve{\bf J}(i\omega+0)
+\frac{1}{\omega^2}\left[-\alpha\dot{\bf X}(0)
+\varepsilon_0\dot{\bf E}(0)\right]\nonumber\\
&&\rule{-6cm}{0cm}-\frac{i}{\omega}\left[-\alpha{\bf X}(0)
+\varepsilon_0{\bf E}(0)\right],\label{4.15}
\end{eqnarray}
where we note that the frequency argument in the susceptibility in front of
$\breve{\bf E}$ differs from that in the corresponding term in
(\ref{4.14}). Adding the right-hand sides of (\ref{4.14})-(\ref{4.15}) we
see that all terms depending on the operators at $t=0$ drop out. The
resulting equality can be used to evaluate the right-hand side of
(\ref{4.13}), if the susceptibility in front of $\breve{\bf E}$ in
(\ref{4.15}) is changed to $\bar{\chi}(-i\omega+0)$ by hand. The correction
term that is brought about in this way, can be rewritten by means of
(\ref{4.5}). Finally, we arrive at the identity:
\begin{equation}
-\alpha{\bf X}^{(+)}({\bf r},\omega)
-\varepsilon_0\chi({\bf r},\omega){\bf E}^{(+)}({\bf r},\omega)=
\frac{i}{\omega}{\bf J}({\bf r},\omega). \label{4.16}
\end{equation}
Hence, the noise-current density ${\bf J}({\bf r},\omega)$ can also be
found as the difference of the positive-frequency Fourier component of the
polarization density $-\alpha({\bf r}){\bf X}({\bf r},t)$ and
$\varepsilon_0\chi({\bf r},\omega)$ times the positive-frequency Fourier
component of the electric field ${\bf E}({\bf r},t)$, apart from a trivial
factor $i/\omega$. Hence, the noise-current density is due to a noise term
in the polarization density, as has been noted for the homogeneous
damped-polariton model \cite{HB92b}, and in the context of the
phenomenological quantization scheme \cite{DKW98,SSG00,KSW01}. In the
present model the noise-current density is a specific linear combination of
the canonical variables, as we shall see below.\\

\section{Evaluation of the noise-current density\label{sectionevaluation}}
The expression (\ref{4.12}) for the noise-current density ${\bf J}({\bf
r},\omega)$ is rather formal. It depends on the forward and the backward
Laplace transforms $\bar{\bf J}$ and $\breve{\bf J}$. In section
\ref{sectionlaplace} these have been given in terms of the canonical
variables. By using (\ref{3.15}) and (\ref{3.21}), we are able to express
${\bf J}({\bf r},\omega)$ in the canonical variables as well.

We start by evaluating the first two contributions at the right-hand side
of (\ref{4.12}). After substitution of (\ref{3.15}) and (\ref{3.21}) and
adding the two contributions, several terms are found to drop out. The
remaining terms are closely related. As before, we shall write
$\bar{\chi}({\bf r},-i\omega+0)$ as $\chi({\bf r},\omega)$, and,
correspondingly, $\bar{\chi}({\bf r},i\omega+0)$ as its complex conjugate
$\chi^{\ast}({\bf r},\omega)$. Furthermore, the imaginary part of
$\chi({\bf r},\omega)$ will be written as $\chi_i({\bf
r},\omega)$. Using these notations, we find from the first two terms of
(\ref{4.12}):
\begin{widetext}
\begin{eqnarray}
\frac{1}{2\pi}
\left[\bar{\bf J}({\bf r},-i\omega+0)
+\breve{\bf J}({\bf r},i\omega+0)\right]&=&
-\frac{\varepsilon_0}{\pi}\, \omega\, \chi_i({\bf r},\omega)\, {\bf A}({\bf r},0)
+\frac{i\varepsilon_0\rho}{\pi\alpha}\, \omega^2\, \chi_i({\bf r},\omega)\,
{\bf X}({\bf r},0)
-\frac{\varepsilon_0}{\pi\alpha}\,\omega\,\chi_i({\bf r},\omega)\,{\bf P}({\bf
r},0)\nonumber\\
&&\rule{-2cm}{0cm}+\frac{i\varepsilon_0}{2\pi\alpha}\,\omega\,\chi({\bf
r},\omega)\int_0^{\infty}d\omega'\, 
\frac{v_{\omega'}}{{\omega'}^2-(\omega+i0)^2}
\left[{\omega'}^2 \, 
{\bf Y}_{\omega'}({\bf r},0)+\frac{i\omega}{\rho}{\bf
Q}_{\omega'}({\bf r},0)\right]\nonumber\\
&&\rule{-2cm}{0cm}-\frac{i\varepsilon_0}{2\pi\alpha}\,\omega\,
\chi^{\ast}({\bf r},\omega)\int_0^{\infty}d\omega'\, 
\frac{v_{\omega'}}{{\omega'}^2-(\omega-i0)^2}
\left[{\omega'}^2\,  
{\bf Y}_{\omega'}({\bf r},0)+\frac{i\omega}{\rho}{\bf
Q}_{\omega'}({\bf r},0)\right],\label{5.1}
\end{eqnarray}
where all canonical variables are taken at $t=0$. It should be noted that
the two integral terms are not the hermitian conjugates of each other,
since the signs of the terms with ${\bf Q}_{\omega'}$ do not match.

In the integral term of (\ref{4.12}) we have to insert (\ref{3.21}) for
$p=i\omega+0$. The term with the spatial derivatives of the vector
potential does not drop out now, as it did in (\ref{5.1}). It can be
evaluated by a partial integration in ${\bf r}'$, which leads to an
expression with derivatives acting on the Green function. Evaluating these
by using (\ref{4.3}), we arrive at the following two terms:
\begin{equation}
-\frac{\varepsilon_0}{\pi c^2}\,\omega^3\,\chi_i({\bf r},\omega)
\int d{\bf r}'\, \varepsilon^{\ast}({\bf r}',\omega)
\, \bfsfG^{\ast}({\bf r},{\bf r}',\omega)\cdot{\bf A}({\bf r}',0)
+\frac{\varepsilon_0}{\pi}\,\omega\,\chi_i({\bf r},\omega)\,{\bf A}({\bf r},0)
.\label{5.2}
\end{equation}
Here we introduced the notation $\bfsfG({\bf r},{\bf
r}',\omega)=\bar{\bfsfG}({\bf r},{\bf r}',-i\omega+0)$ (and hence
$\bfsfG^{\ast}({\bf r},{\bf r}',\omega)=\bar{\bfsfG}({\bf r},{\bf
r}',i\omega+0)$ as well), in analogy to the notations for $\varepsilon$ and
$\chi$. The final term in (\ref{5.2}) cancels the first term in
(\ref{5.1}). Part of the integral term in (\ref{5.2}) (namely, with
$\chi^{\ast}$ instead of $\varepsilon^{\ast}$) drops out as well, when the
contribution from the second term in (\ref{3.21}) is taken into account.

Collecting all terms, we arrive at the following result for the
noise-current density:
\begin{eqnarray}
{\bf J}({\bf r},\omega)&=&\int d{\bf r}'\, \left\{
\bfsfc_A({\bf r},{\bf r}',\omega)\cdot{\bf A}({\bf r}',0)
+\bfsfc_{\Pi}({\bf r},{\bf r}',\omega)\cdot{\bm \Pi}({\bf r}',0)
+\bfsfc_X({\bf r},{\bf r}',\omega)\cdot{\bf X}({\bf r}',0)
+\bfsfc_P({\bf r},{\bf r}',\omega)\cdot{\bf P}({\bf r}',0)\right.\nonumber\\
&&\left.+\int_0^{\infty}d\omega'\, 
\bfsfc_{YQ}({\bf r},{\bf r}',\omega,\omega')\cdot
\left[{\omega'}^2\, 
{\bf Y}_{\omega'}({\bf r}',0)+\frac{i\omega}{\rho'}
{\bf Q}_{\omega'}({\bf r}',0)\right]\right\}.\label{5.3}
\end{eqnarray}
The (tensorial) coefficients have the following form
\begin{subequations}
\label{5.4}
\begin{eqnarray}
\bfsfc_A({\bf r},{\bf r}',\omega)&=&-\frac{\varepsilon_0}{\pi c^2}\,\omega^3
\chi_i({\bf r},\omega)\, 
\bfsfG_{{\rm T}'}^{\ast}({\bf r},{\bf r}',\omega),\label{5.4a}\\
\bfsfc_{\Pi}({\bf r},{\bf r}',\omega)&=&-\frac{i}{\pi c^2}\,\omega^2\,
\chi_i({\bf r},\omega)\, 
\bfsfG_{{\rm T}'}^{\ast}({\bf r},{\bf r}',\omega),\label{5.4b}\\
\bfsfc_X({\bf r},{\bf r}',\omega)
&=&\frac{i\varepsilon_0\rho}{\pi\alpha}\,\omega^2\,
\chi_i({\bf r},\omega)\, \bfsfI\, \delta({\bf r}-{\bf r}')
-\frac{i\alpha'}{\pi c^2}\,\omega^2\,\chi_i({\bf r},\omega)\,
\bfsfG_{{\rm T}'}^{\ast}({\bf r},{\bf r}',\omega)\nonumber\\
&&-\frac{i\varepsilon_0\rho'}{\pi c^2\alpha'}\,
\omega^4\,\chi_i({\bf r},\omega)\,\chi^{\ast}({\bf r}',\omega)\,
\bfsfG^{\ast}({\bf r},{\bf r}',\omega),\rule{1cm}{0cm}\label{5.4c}\\
\bfsfc_P({\bf r},{\bf r}',\omega)
&=&-\frac{\varepsilon_0}{\pi\alpha}\,\omega\,\chi_i({\bf r},\omega)
\, \bfsfI\, \delta({\bf r}-{\bf r}')
+\frac{\varepsilon_0}{\pi c^2\alpha'}\,\omega^3\,
\chi_i({\bf r},\omega)\,\chi^{\ast}({\bf r}',\omega)\,
\bfsfG^{\ast}({\bf r},{\bf r}',\omega),\label{5.4d}\\
\bfsfc_{YQ}({\bf r},{\bf r}',\omega,\omega')&=&
-\frac{\varepsilon_0}{\pi\alpha}\;{\rm Im}\left[
\frac{\omega\, v_{\omega'}}{{\omega'}^2-(\omega+i0)^2}\,
\chi({\bf r},\omega)\right]\, \bfsfI\, \delta({\bf r}-{\bf r}')\nonumber\\
&&+\frac{\varepsilon_0}{\pi c^2\alpha'}
\frac{\omega^3\, v'_{\omega'}}{{\omega'}^2-(\omega-i0)^2}\,
\chi_i({\bf r},\omega)\,\chi^{\ast}({\bf r}',\omega)\,
\bfsfG^{\ast}({\bf r},{\bf r}',\omega).\label{5.4e}
\end{eqnarray}
\end{subequations}
\end{widetext}
In the first three formulas the complex conjugate of the Green function
$\bfsfG_{{\rm T}'}({\bf r},{\bf r}',\omega)$ appears. It is the transverse
part of $\bfsfG({\bf r},{\bf r}',\omega)$ with respect to ${\bf r}'$, which
is defined by the convolution $\int d{\bf r}''\, \bfsfG({\bf r},{\bf
r}'',\omega)\cdot {\bm \delta}_{\rm T}({\bf r}''-{\bf r}')$. In the last
term of (\ref{5.4e}) the symbol $v'_{\omega'}$ denotes the bath coupling
parameter at the position ${\bf r}'$ (and the frequency
$\omega'$). Furthermore, $\alpha'$ and $\rho'$ stand for $\alpha({\bf r}')$
and $\rho({\bf r}')$, respectively, as before.

The coefficients (\ref{5.4}) can be interpreted as commutators. In fact,
from (\ref{2.4}) we infer
\begin{subequations}
\label{5.5}
\begin{eqnarray}
\frac{i}{\hbarit}\left[{\bf J}({\bf
r},\omega),{\bm \Pi}({\bf r}',0)\right]&=&
-\bfsfc_A({\bf r},{\bf r}',\omega),\label{5.5a}\\
\frac{i}{\hbarit}\left[{\bf J}({\bf
r},\omega),{\bf A}({\bf r}',0)\right]&=&
\bfsfc_{\Pi}({\bf r},{\bf r}',\omega),\label{5.5b}\\
\frac{i}{\hbarit}\left[{\bf J}({\bf
r},\omega),{\bf X}({\bf r}',0)\right]&=&
\bfsfc_{P}({\bf r},{\bf r}',\omega),\label{5.5c}
\end{eqnarray}
\end{subequations}
and analogous relations for the other coefficients.

It should be noted that the coefficients (\ref{5.4a})-(\ref{5.4d}) are all
proportional to the imaginary part $\chi_i({\bf r},\omega)$ of the
susceptibility. Furthermore, the coefficient (\ref{5.4e}) is proportional
to the bath coupling parameter $v_{\omega}$. In the absence of absorption
the dielectric is not coupled to a bath, so that $v_{\omega}$ vanishes. As
(\ref{3.7}) shows, the imaginary part of the susceptibility vanishes in
that case as well, at least for all frequencies that are
off-resonance. Hence, all coefficients (\ref{5.4}) are zero in this case,
so that the noise-current density itself disappears. Clearly, the present
formalism loses its meaning for a non-absorptive dielectric. 

As the noise-current density is fully known now in terms of the canonical
variables, we can proceed and derive its properties. This will be the
subject of the next section.\\

\section{Properties of the noise-current density\label{sectionproperties}}
In this section we will determine a few of the properties of the
noise-current density ${\bf J}({\bf r},\omega)$. In particular, we will
focus on its commutation relations.

We start by considering the commutator of ${\bf J}({\bf r},\omega)$ with
the Hamiltonian (\ref{2.3}). To evaluate this commutator, we might use the
expressions (\ref{5.3})-(\ref{5.4}), employ the canonical commutation
relations (\ref{2.4}) and evaluate all contributions in a systematic
way. Owing to the complexity of (\ref{5.4}), this is a rather tedious
task. A more convenient way to obtain the commutator is to use the
expression (\ref{4.12}) for ${\bf J}({\bf r},\omega)$ in terms of $\bar{\bf
J}({\bf r},-i\omega+0)$ and $\breve{\bf J}({\bf r},i\omega+0)$. The
commutators of the latter with the Hamiltonian can be found without
difficulty. In fact, one gets by evaluating the commutators of (\ref{3.15})
and (\ref{3.21}) with (\ref{2.3}), or more straightforwardly, by employing
the equations of motion (\ref{2.6}):
\begin{eqnarray}
\frac{i}{\hbarit}[H,\bar{\bf J}({\bf r},p)]&=&p\bar{\bf J}({\bf r},p)
\nonumber\\
&&\rule{-3cm}{0cm}
+\frac{1}{\mu_0 p} \left\{ {\bm \nabla}\times[{\bm \nabla}
\times{\bf E}({\bf r},0)]
+\frac{p^2}{c^2}\bar{\varepsilon}({\bf r},p){\bf E}({\bf r},0)\right\},
\label{6.1}
\end{eqnarray}
and
\begin{eqnarray}
\frac{i}{\hbarit}[H,\breve{\bf J}({\bf r},p)]&=&-p\breve{\bf J}({\bf r},p)
\nonumber\\
&&\rule{-3cm}{0cm}
+\frac{1}{\mu_0 p}\left\{ {\bm \nabla}\times[{\bm \nabla}
\times{\bf E}({\bf r},0)]
+\frac{p^2}{c^2}\bar{\varepsilon}({\bf r},p){\bf E}({\bf r},0)\right\},
\label{6.2}
\end{eqnarray}
where the electric field is taken at time $t=0$.  Using these
expressions, we find as the contribution from the first two terms in
(\ref{4.12}) to the commutator $(i/\hbarit) [H,{\bf J}({\bf r},\omega)]$:
\begin{eqnarray}
-\frac{i\omega}{2\pi}\left[\bar{\bf J}({\bf r},-i\omega+0)
+\breve{\bf J}({\bf r},i\omega+0)\right]\nonumber\\
\rule{2cm}{0cm}
+\frac{\varepsilon_0}{\pi}\omega\chi_i({\bf r},\omega){\bf E}({\bf r},0). 
\label{6.3}
\end{eqnarray}
Furthermore, the last term in (\ref{4.12}) contributes:
\begin{eqnarray}
\frac{1}{\pi c^2}\omega^3\chi_i({\bf r},\omega)
\int d{\bf r}'\, \bar{\bfsfG}({\bf r},{\bf r}',i\omega +0)\cdot
\breve{\bf J}({\bf r}',i\omega+0)\nonumber\\
-\frac{\varepsilon_0}{\pi} \omega
\chi_i({\bf r},\omega) {\bf E} ({\bf r},0). \label{6.4}
\end{eqnarray}
On adding the two contributions, we see that the terms depending on ${\bf
E}({\bf r},0)$ cancel. The remaining terms are proportional to ${\bf J}({\bf
r},\omega)$, so that we arrive at the simple result
\begin{equation}
\frac{i}{\hbarit}\left[H,{\bf J}({\bf r},\omega)\right]=
-i\omega{\bf J}({\bf r},\omega). \label{6.5}
\end{equation}

To understand how this commutation property comes about, it is useful to
give a somewhat more formal derivation of the commutator. To that end we
start by remarking that for an arbitrary operator $\Omega(t)$ the
commutator of the Hamiltonian with its Laplace transform $\bar{\Omega}(p)$
follows directly by Laplace-transforming the equation of motion in the
Heisenberg picture:
\begin{equation}
\frac{i}{\hbarit}[H,\bar{\Omega}(p)]=p\bar{\Omega}(p)-\Omega(0).
\label{6.6}
\end{equation}
Writing the analogous equation for the backward Laplace transform:
\begin{equation}
\frac{i}{\hbarit}[H,\breve{\Omega}(p)]=-p\breve{\Omega}(p)+\Omega(0),
\label{6.7}
\end{equation}
and adding the two equations after substitution of the appropriate
arguments $p$, we get
\begin{eqnarray}
\frac{i}{\hbarit}\left[H,\{
\bar{\Omega}(-i\omega+0)+\breve{\Omega}(i\omega+0)\}\right]\nonumber\\
=-i\omega\{\bar{\Omega}(-i\omega+0)+\breve{\Omega}(i\omega+0)\}.
\label{6.8}
\end{eqnarray}
In particular, one gets for $\Omega(t)={\bf E}({\bf r},t)$ by comparison
with (\ref{4.10}):
\begin{equation}
\frac{i}{\hbarit}\left[H,{\bf E}^{(+)}({\bf r},\omega)\right]=
-i\omega{\bf E}^{(+)}({\bf r},\omega). \label{6.9}
\end{equation}
Of course, this could not be otherwise: if it did not hold, the time dependence
in (\ref{4.8}) would be compromised. By invoking the definition (\ref{4.11}) of
the noise-current density in terms of the positive-frequency Fourier component
of the electric field, it is immediately clear now that ${\bf J}({\bf
r},\omega)$ must satisfy a commutation relation of the same form, which is
indeed what we got in (\ref{6.5}).

It should be noted that the commutator expression in (\ref{6.1}) contains
an additional term that differs from that in (\ref{6.6}). The reason for
this discrepancy is that $\bar{\bf J}({\bf r},p)$ has not been defined as
the Laplace transform of some operator ${\bf J}({\bf r},t)$, as we noticed
already in Sec.\ \ref{sectionlaplace}. Similar remarks apply to
$\breve{\bf J}({\bf r},p)$.

Let us now turn our attention to the commutator of ${\bf J}({\bf
r},\omega)$ with its hermitian conjugate at a different position and
frequency. In view of the general form (\ref{4.12}) it is convenient to
start by calculating the three commutators involving $\bar{\bf J}$ and
$\breve{\bf J}$. These follow by substitution of (\ref{3.15}) and
(\ref{3.21}) and use of the canonical commutation relations
(\ref{2.4}). The results are given in (\ref{B.1}) and (\ref{B.3}) of
Appendix \ref{appendixb}. As shown there, these commutators of $\bar{\bf
J}$ and $\breve{\bf J}$ can be used to prove the commutation relation:
\begin{eqnarray}
\left[{\bf J}({\bf r},\omega),
\left[{\bf J}({\bf r}',\omega')\right]^{\dagger}\right]&&\nonumber\\
&&\rule{-3cm}{0cm}=\frac{\varepsilon_0\hbarit}{\pi}\, \omega^2\, 
\chi_i({\bf r},\omega)\, 
\delta(\omega-\omega')\, \bfsfI\, \delta({\bf r}-{\bf r}'). \label{6.10}
\end{eqnarray}
In an analogous fashion one may evaluate the commutator of the
noise-current density with its counterpart for different arguments. It is
found that this commutator vanishes:
\begin{equation}
\left[{\bf J}({\bf r},\omega),{\bf J}({\bf r}',\omega')\right]=0 .
\label{6.11}
\end{equation}
As shown in appendix \ref{appendixb}, the commutators (\ref{6.10}) and
(\ref{6.11}) appear as the results of calculations in which several terms
cancel one another. In fact, all nonlocal terms involving transverse delta
functions and Green functions drop out. The final answers show that ${\bf
J}({\bf r},\omega)$ is a strictly local operator in its space variable: for
all ${\bf r}'\neq {\bf r}$ it commutes both with ${\bf J}({\bf r}',\omega)$
and with the hermitian conjugate of the latter. Moreover, the right-hand
sides of (\ref{6.10})-(\ref{6.11}) show that the noise-current density is
local in the frequency as well: the commutators vanish for $\omega\neq
\omega'$. As a final comment we note that the commutator (\ref{6.10}) is
proportional to the imaginary part of the susceptibility. Hence, it
vanishes if there is no absorption. As we have seen above, the
noise-current density itself vanishes in that case, so that (\ref{6.10})
becomes a trivial identity.

As demonstrated above, the commutator properties (\ref{6.5}), (\ref{6.10})
and (\ref{6.11}) follow from the dynamics of the damped-polariton
model. These commutation relations are the same as the postulated relations
of the noise-current density in the phenomenological quantization scheme
\cite{GW95, MLBJ95,GW96a,ML96,GW96b,DKW98,SKW98}. Evidently, the status of
the commutation relations is rather different in both schemes. Our results
provide a justification for the postulates in the phenomenological theory.

The collection of operators ${\bf J}({\bf r},\omega)$ possesses another
convenient property: together with their hermitian conjugates they form a
complete basis set for the canonical variables of the model. It means that
each of these variables can be written as a linear combination of
the operators from the basis. To prove this statement, we may argue as
follows. Let us tentatively write the vector potential as
\begin{equation}
{\bf A}({\bf r},0)=\int d{\bf r}' \int_0^{\infty} d \omega\, 
{\bf J}({\bf r}',\omega)\cdot\bfsff_{A}({\bf r}',{\bf r},\omega)
 +\text{h.c.},\label{6.12}
\end{equation} 
with as yet unknown tensorial coefficients $\bfsff_{A}$. Taking the commutator
of both sides with $[{\bf J}({\bf r}'',\omega')]^{\dagger}$ we find from
(\ref{5.5b}), (\ref{6.10}) and (\ref{6.11}):
\begin{equation}
\bfsff_A({\bf r},{\bf r}',\omega)=
\frac{i\pi}{\varepsilon_0\omega^2\chi_i({\bf r},\omega)}\, 
\bfsfc^{\ast}_{\Pi}({\bf r},{\bf r}',\omega), \label{6.13}
\end{equation}
so that we get:
\begin{eqnarray}
{\bf A}({\bf r},0)=\frac{i\pi}{\varepsilon_0}\int d{\bf r}' \int_0^{\infty}
d \omega\, \frac{1}{\omega^2 \chi_i({\bf r}',\omega)}\nonumber\\
\times{\bf J}({\bf r}',\omega)\cdot\bfsfc_{\Pi}^{\ast}({\bf r}',{\bf r},\omega) 
+ \text{h.c.}\; .  \label{6.14}
\end{eqnarray}
To really establish the validity of this equality, which we found by merely
assuming the general form (\ref{6.12}), we insert (\ref{5.3}) in the
right-hand side, which then becomes a linear combination of the canonical
variables. Upon evaluating the resulting integrals with the techniques of
Appendices \ref{appendixa} and \ref{appendixb}, we indeed find that only
the term with the vector potential survives, and that the left-hand side is
recovered.

\begin{widetext}
Two other examples of identities, which may be checked in an analogous way,
are:
\begin{subequations}
\label{6.15}
\begin{eqnarray}
{\bm \Pi}({\bf r},0)&=&-\frac{i\pi}{\varepsilon_0}\int d{\bf r}' \int_0^{\infty}
d \omega\, \frac{1}{\omega^2 \chi_i({\bf r}',\omega)}\, 
{\bf J}({\bf r}',\omega)\cdot\bfsfc_{A}^{\ast}({\bf r}',{\bf r},\omega) +
\text{h.c.},\label{6.15b}\\
{\bf X}({\bf r},0)&=&\frac{i\pi}{\varepsilon_0}\int d{\bf r}' \int_0^{\infty}
d \omega\, \frac{1}{\omega^2 \chi_i({\bf r}',\omega)}\, 
{\bf J}({\bf r}',\omega)\cdot\bfsfc_{P}^{\ast}({\bf r}',{\bf r},\omega) +
\text{h.c.}\; .\label{6.15c}
\end{eqnarray}
\end{subequations}
Similar identities are found to be valid for the canonical
variables ${\bf P}$, ${\bf Y}_{\omega}$ and ${\bf Q}_{\omega}$. Since all
canonical variables can thus be expressed in terms of ${\bf J}$ and its
hermitian conjugate, these operators must form a complete basis, as we set
out to prove.

The completeness of the set of operators ${\bf J}({\bf r},\omega)$ and
$[{\bf J}({\bf r},\omega)]^{\dagger}$, and their properties (\ref{6.5}),
(\ref{6.10}) and (\ref{6.11}) imply that the noise-current density is
proportional to the diagonalizing operator of the system Hamiltonian
(\ref{2.3}). In fact, we may write:
\begin{equation}
H=\frac{\pi}{\varepsilon_0}\int d{\bf r}\int_0^{\infty} d\omega 
\frac{1}{\omega\, \chi_i({\bf r},\omega)} \, 
\left[ {\bf J}({\bf r},\omega)\right]^{\dagger}\cdot {\bf J}({\bf r},\omega). 
\label{6.16}
\end{equation}
We have checked that (\ref{2.3}) can be recovered from (\ref{6.16}). This
is accomplished by substituting (\ref{5.3}) with (\ref{5.4}) in
(\ref{6.16}) and evaluating the resulting expression in terms of the
canonical variables. A few details of this calculation are given in
Appendix C. It should be noted that the two expressions for the Hamiltonian
do not agree completely: they differ by a $c$-number, which corresponds to
a zero-point energy.

Now that we have succeeded in obtaining the diagonalizing operators of our
model, we can determine the full time dependence of the vector potential, the
electric field, the polarization density, or any of the dynamic variables that
we have considered above. For example, the vector potential at time $t$ follows
from (\ref{6.14}) by substituting the time-dependent noise-current density
$e^{-i\omega t}{\bf J}({\bf r}',\omega)$ in the integrand. Inserting the
expression (\ref{5.4b}) for $\bfsfc_{\Pi}$, and using (\ref{4.2}) we get
\begin{equation}
{\bf A}({\bf r},t)=-\mu_0\int d{\bf r}'\int_0^{\infty}d\omega\, 
e^{-i\omega t}\, 
\bfsfG_{\rm T}({\bf r},{\bf r}',\omega)\cdot{\bf J}({\bf r}',\omega) + 
\text{h.c.}\; .
\label{6.17}
\end{equation}
In the phenomenological quantization scheme an integral representation of
the same form shows up \cite{DKW98}. However, in that theory the
noise-current density ${\bf J}({\bf r},\omega)$ is a formal operator. In
the present model we have an explicit expression for ${\bf J}$ at our
disposal. In fact, by substituting (\ref{5.3}) we may evaluate the
right-hand side of (\ref{6.17}) in terms of the canonical variables at
$t=0$. The results are presented in Appendix \ref{appendixd}. As shown
there, the vector potential gets a simple form when enough time has passed
for transients to die out. In that long-time limit, it reduces to a
linear combination of bath operators only:
\begin{equation}
{\bf A}({\bf r},t)\simeq 
\frac{1}{2c^2}\int d{\bf r}'\, \frac{1}{\alpha'} \int_0^{\infty}
d \omega\, e^{-i\omega t}\, v'(\omega)\, \chi({\bf r}',\omega)\, 
\bfsfG_{\rm T}({\bf r},{\bf r}',\omega)\cdot
\left[\omega^2\, {\bf Y}_{\omega}({\bf r}',0)+\frac{i\omega}{\rho'}\, 
{\bf Q}_{\omega}({\bf r}',0)\right] + \text{h.c.}\; . \label{6.18}
\end{equation} 
This expression has the same form as (\ref{6.17}), with the noise-current
operator replaced by:
\begin{equation}
{\bf J}_l({\bf r},\omega)=-\frac{\varepsilon_0}{2\alpha}\, 
v(\omega)\, \chi({\bf r},\omega)\, 
\left[\omega^2\,{\bf Y}_{\omega}({\bf r},0)+\frac{i\omega}{\rho}\, 
{\bf Q}_{\omega}({\bf r},0)\right]. \label{6.19}
\end{equation} 
The combination between square brackets is proportional to the annihilation
operator of the bath harmonic oscillators at the chosen position and
frequency. Indeed, ${\bf J}_l$ satisfies the same standard commutation relations
(\ref{6.10})-(\ref{6.11}) as ${\bf J}$. The vector potential thus depends on the
bath annihilation and creation operators only, when all transients have died
out. This result for the long-time limit is the generalization of a similar
finding for the homogeneous damped-polariton model, which we discussed before
\cite{WS01}.

The time-dependent electric field ${\bf E}({\bf r},t)$ could be found in
principle by separately evaluating its transverse part $-{\bm \Pi}({\bf
r},t)/\varepsilon_0$ from (\ref{6.15b}) and its longitudinal part
$[\alpha{\bf X}({\bf r},t)]_L/\varepsilon_0$ from (\ref{6.15c}), and adding
the two contributions. However, a simpler way to obtain ${\bf E}({\bf
r},t)$ is to insert the solution of (\ref{4.11}) into the general form
(\ref{4.8}). In this way we get
\begin{equation}
{\bf E}({\bf r},t)=-i\mu_0 \int d{\bf r}' \int_0^{\infty}
d \omega\, e^{-i\omega t}\,\omega \, \bfsfG({\bf r},{\bf r}',\omega)
\cdot{\bf J}({\bf r}',\omega)
+\text{h.c.}\; . \label{6.20}
\end{equation}
In Appendix \ref{appendixd}, it is shown that in the long-time limit the
electric field is given by an expression of the same form as (\ref{6.20}),
with ${\bf J}$ replaced by ${\bf J}_l$. By using the commutation relations
(\ref{6.10})-(\ref{6.11}), which are valid for ${\bf J}_l$ as well, one may
show that the commutator of the electric field and the vector potential in
the long-time limit has the standard form (\ref{2.5a}).

As a final example, we consider the time-dependent polarization density
$-\alpha {\bf X}({\bf r},t)$. It follows from (\ref{6.15c}) as:
\begin{equation}
-\alpha{\bf X}({\bf r},t)=-\frac{i\pi\alpha}{\varepsilon_0}\int d{\bf r}' 
\int_0^{\infty}
d \omega\, \frac{1}{\omega^2 \chi_i({\bf r}',\omega)}\, e^{-i\omega t}\,
{\bf J}({\bf r}',\omega)\cdot\bfsfc_{P}^{\ast}({\bf r}',{\bf r},\omega) +
\text{h.c.}\; .\label{6.21}
\end{equation}
Substituting (\ref{5.4d}) we get:
\begin{equation}
-\alpha{\bf X}({\bf r},t)=-\frac{i}{c^2}\int d{\bf r}'\int_0^{\infty}d\omega\, 
e^{-i\omega t}\, \omega \, \chi({\bf r},\omega)\, 
\bfsfG({\bf r},{\bf r}',\omega)\cdot{\bf J}({\bf r}',\omega)
+i\int_0^{\infty} d\omega\, e^{-i\omega t}\, \frac{1}{\omega}\,
{\bf J}({\bf r},\omega)+\text{h.c.}\; , \label{6.22}
\end{equation}
\end{widetext}
where we used (\ref{4.2}).  This form for the time-dependent polarization
density shows that it is the sum of a term involving the properties of the
medium through the susceptibility and a term which is determined by the
noise-current density only. In fact, this is consistent with (\ref{4.16}),
which was written in terms of the positive-frequency Fourier
components. Indeed, the integrand in the first term at the right-hand side
of (\ref{6.22}) is proportional to the positive-frequency part of the
electric field, as we have seen in (\ref{6.20}). In Appendix
\ref{appendixd} the time-dependent polarization density is evaluated in
terms of the canonical variables at $t=0$. Furthermore, it is shown there
that the long-time limit of $-\alpha {\bf X}({\bf r},t)$ follows from
(\ref{6.22}) by replacing ${\bf J}$ by ${\bf J}_l$, as was found above for
the vector potential and the electric field.

The expressions (\ref{6.17}), (\ref{6.20}) and (\ref{6.22}) give the
complete time dependence of the vector potential, the electric field and
the polarization density in the inhomogeneous damped-polariton model.  For
the special case of a homogeneous medium, the expressions reduce to those
given in \cite{HB92b}. As we have seen, the implicit dependence on the
canonical variables at $t=0$ can be made explicit by substitution of
the noise-current density in the form of (\ref{5.3}).

Now that we have found in (\ref{6.17}) and (\ref{6.20}) the explicit time
dependence of the field operators ${\bf A}$ and ${\bf E}$ in terms of the
noise-current density ${\bf J}$, we can determine the commutators $[{\bf
E}({\bf r},t),{\bf A}({\bf r'},t)]$ and $[{\bf E}({\bf r},t),{\bf B}({\bf
r'},t)]$ for arbitrary $t$. Actually, for $t=0$ we have already determined
these in (\ref{2.5}) and the fact that these commutators are
medium-independent almost directly followed from the standard commutation
relations (\ref{2.4}). Now that we have integrated out the dynamics of the
material variables, the expressions (\ref{6.17}) and (\ref{6.20}) clearly
both depend on the medium through the Green function $\bfsfG({\bf r},{\bf
r}',\omega)$ and the noise-current density ${\bf J}({\bf
r}',\omega)$. However, since no approximations were made in order to obtain
the time dependence of the field operators, their commutators should still
be medium-independent, and equal to those at $t=0$. With the use of
(\ref{4.1})-(\ref{4.3}) and the Green function sum rule (\ref{A.4}) one can
can verify that the commutators $[{\bf E}({\bf r},t),{\bf A}({\bf r'},t)]$
and $[{\bf E}({\bf r},t),{\bf B}({\bf r'},t)]$ indeed have the
medium-independent values of (\ref{2.5}). A medium-independent commutator
$[{\bf E}({\bf r},t),{\bf B}({\bf r'},t)]$ was also found in the
phenomenological scheme \cite{SKW98}, which was the principal argument in
showing that the phenomenological scheme is consistent with (although not
founded on) quantum electrodynamics. Finally, it may be remarked that in
our theory the commutators $[{\bf P}({\bf r},t),{\bf X}({\bf r'},t)]$ and
$[{\bf Q}_{\omega}({\bf r},t),{\bf Y}_{\omega'}({\bf r'},t)]$ are also
medium-independent; these commutators have no counterparts in the
phenomenological theory.\\

\section{Conclusion and discussion\label{sectionconclusion}}
By solving the inhomogeneous damped-polariton model we have established a
rigorous basis for the phenomenological quantization procedure, which has been
used to describe quantum phenomena in linear lossy dielectrics with great
success. Up to now such a firm basis was available for homogeneous dielectrics
only, through the pioneering work of Huttner and Barnett
\cite{HB92a,HB92b}. 

As a tool in our treatment we have used forward and backward Laplace
transformations. With the help of these we solved the equations of motion
for the canonical variables. The Laplace transforms of the electric field
were shown to obey wave equations with source terms that could be expressed
in terms of the canonical variables at time $t=0$. Upon introducing the
Green function of these wave equations we were able to derive an expression
for the positive-frequency Fourier component of the time-dependent electric
field. The latter was found to satisfy a wave equation with a
frequency-dependent source term that could be interpreted as a
noise-current density for the inhomogeneous damped-polariton
model. Explicit expressions for this noise-current density in terms of the
canonical variables of the system have been derived. By establishing its
algebraic properties we could prove that it is proportional to the
diagonalizing operator of the model. Once we have shown this, the
time-dependence of all relevant operators can be determined. As an
illustration we gave the time-dependent expressions for the vector
potential, the electric field and the polarization density.

In order to show the internal consistency of our results, we have derived
and employed several frequency sum rules for the tensorial Green function
and for the susceptibility, namely (\ref{A.4})-(\ref{A.5}), (\ref{C.6}) and
(\ref{C.8})-(\ref{C.9}). The outcomes solely depend on the high-frequency
asymptotic behavior of the Green function and the susceptibility. In the
present model this behavior is determined by the values of the (local)
parameters $\alpha$, $\rho$, $\omega_0$ and $v_{\omega}$. The Green
function sum rules are generalizations of `velocity sum rules' that have
been derived for homogeneous dielectrics \cite{HB92b,WS01}.

We have expressed all field operators in terms of the noise-current density
operators ${\bf J}({\bf r},\omega)$, which also diagonalize the
Hamiltonian. These operators and their hermitian conjugates were proved to
be local both in position and in frequency: any pair of them commute when
taken at different positions and/or different frequencies. The locality in
position is not self-evident {\em a priori}, as some of the canonical
variables of the model, namely ${\bf A}$ and ${\bm \Pi}$, satisfy a
commutation relation (\ref{2.4a}) with a non-local transverse delta
function. The positive-frequency Fourier component ${\bf E}^{(+)}({\bf
r},\omega)$ of the electric field is non-local in space as well: it does
not commute with its hermitian conjugate at a position ${\bf r}'$ (and at
the same frequency $\omega$). The locality of ${\bf J}({\bf r},\omega)$
with respect to the frequency is connected to the validity of a generalized
optical theorem (\ref{B.7}) for the Green function. When the independent
frequency variables in this theorem are chosen to be equal, it reduces to
the standard form of the optical theorem \cite{DKW98} .

The diagonalizing operators are not unique. For example, if one breaks up
the noise-current density in terms of its canonical elements according to
(\ref{5.3}), then for long times after the initial time $t=0$ one finds
that the field operators are determined only by the initial bath operators,
since time-dependent coefficients of other canonical variables all decay
exponentially fast. If only long times are considered, ${\bf J}({\bf
r},\omega)$ can be taken to be proportional to the initial annihilation
operator of the bath harmonic oscillator at position ${\bf r}$ and
frequency $\omega$. We stressed this point in \cite{WS01} for homogeneous
dielectrics. Other diagonalizing operators can be constructed by
transforming the noise-current density ${\bf J}({\bf r},\omega)$ with
arbitrary unitary transformations $\bfsfU({\bf r},{\bf r'},\omega)$, but
these would not have the physical interpretation of noise-current density
operators.

Our solution provides detailed information on the dynamical behavior of
absorptive dielectrics. This information can be used to study dynamical
processes like spontaneous emission of guest atoms in inhomogeneous
media. For instance, transient effects in emission processes, which we
studied before in homogeneous media \cite{WS01}, can now be investigated in
the general inhomogeneous case. Local-field effects, which by their very
nature are brought about by inhomogeneities in the medium, form another
field of interest for which our solution may be helpful.

To obtain our results we have employed a Laplace-transform technique that we
used before \cite{WS01}. An alternative method, which was adopted in
\cite{HB92b}, is based on a diagonalization procedure due to Fano
\cite{F61}. We have been able to carry out the diagonalization of the
inhomogeneous damped-polariton model along those lines as well. Details of
that work will be published elsewhere \cite{SW04}.\\

\section*{Acknowledgments}
We would like to thank dr A.J.\ van Wonderen for many stimulating
discussions.\\

\appendix
\section{Short-time limit of the electric field\label{appendixa}}
In Sec.\ \ref{sectiongreen} the electric-field operator ${\bf E}({\bf
r},t)$ has been found as a Fourier integral (\ref{4.8}), with the
positive-frequency Fourier component ${\bf E}^{(+)}({\bf r},\omega)$ given in
(\ref{4.9}). It contains the source terms $\bar{\bf J}$ and $\breve{\bf
J}$, which have been given in (\ref{3.15}) and (\ref{3.21}) as linear
combinations of the canonical variables at $t=0$. As a check, we shall
verify that the Fourier-integral in (\ref{4.8}) reduces to ${\bf E}({\bf
r},0)$ in the limit $t\rightarrow 0$. As we shall see, the proof will
depend on the validity of a few sum rules for the Green function.

By employing the identity $[\bar{\bf E}({\bf r},-i\omega+0)]^\dagger=
\bar{\bf E}({\bf r},i\omega+0)$, and the corresponding identity for the
backward Laplace transform, one may write the Fourier integral
representing ${\bf E}({\bf r},0)$ as
\begin{equation}
\frac{1}{2\pi}\int_{-\infty}^{\infty}d\omega\, \left[
\bar{\bf E}({\bf r},-i\omega+0)+\breve{\bf E}({\bf r},-i\omega+0)\right].
\label{A.1}
\end{equation}
Upon substituting (\ref{4.4})-(\ref{4.5}), (\ref{3.15}) and (\ref{3.21}),
we find that all terms with ${\bf A}({\bf r}',0)$, ${\bf P}({\bf r}',0)$
and ${\bf Y}_{\omega'}({\bf r}',0)$ cancel. We are left with the following
expression:
\begin{eqnarray}
&&-\frac{i}{\pi}\int_{-\infty}^{\infty}d\omega\int d{\bf r}'\,
\bar{\bfsfG}({\bf r},{\bf r}',-i\omega+0)\cdot\biggl\{\mu_0\omega\,{\bm
\Pi}({\bf r}',0)\nonumber\\
&&+\mu_0\omega[\alpha'{\bf X}({\bf r}',0)]_{{\rm T}'}
+\frac{\rho'}{c^2\alpha'}\omega^3\bar{\chi}({\bf r}',-i\omega+0)\,
{\bf X}({\bf r}',0)\nonumber\\
&&-\frac{1}{c^2\alpha'\rho'}\omega^3\bar{\chi}({\bf r}',-i\omega+0)
\nonumber\\
&&\rule{1cm}{0cm}
\times\int_0^{\infty}d\omega'\,\frac{v'_{\omega'}}{{\omega'}^2-(\omega+i0)^2}\,
{\bf Q}_{\omega'}({\bf r}',0)\biggr\}. \label{A.2}
\end{eqnarray}
This result can be simplified by considering the integrals over
$\omega$. The terms with ${\bm \Pi}$ and $[\alpha{\bf X}]_T$ contain the
integral $\int_{-\infty}^{\infty}d\omega\, \omega\, \bar{\bfsfG}({\bf
r},{\bf r}',-i\omega+0)$. Since the Green function is analytic for $\omega$
in the upper half-plane, we may evaluate the integral by closing the contour
in this half-plane. The Green function satisfies the differential equation
(\ref{4.1}). Now $\bar{\varepsilon}({\bf r},p)=1+\bar{\chi}({\bf r},p)$
tends to 1 for large $p$ in the right half-plane, as follows from the
expression (\ref{3.7}) for $\bar{\chi}$. Hence, the asymptotic form of the
Green function for large $\omega$ in the upper half-plane has the same form as
the free-space Green function, namely:
\begin{equation}
\bar{\bfsfG}({\bf r},{\bf r}',-i\omega+0)\simeq
\frac{c^2}{(\omega+i0)^2}\, \bfsfI\, \delta({\bf r}-{\bf r}').\label{A.3}
\end{equation}
As a consequence, closing the contour yields the identity:
\begin{equation}
\int_{-\infty}^{\infty}d\omega\, \omega\, \bar{\bfsfG}({\bf r},{\bf
r}',-i\omega+0)=-i\pi c^2\, \bfsfI\,\delta({\bf r}-{\bf r}'),\label{A.4}
\end{equation}
which is in fact a sum rule for the Green function \cite{SKW98}.

Likewise, one proves
\begin{eqnarray}
\int_{-\infty}^{\infty}d\omega\, \omega^3\, \bar{\bfsfG}({\bf r},{\bf
r}',-i\omega+0)\bar{\chi}({\bf r}',-i\omega+0)&&\nonumber\\
&&\rule{-3cm}{0cm}
=\frac{i\pi c^2\alpha^2}{\varepsilon_0\rho}\, \bfsfI\,\delta({\bf r}-{\bf r}'),
\label{A.5}
\end{eqnarray}
since the susceptibility $\bar{\chi}({\bf r},-i\omega+0)$ behaves like
$-[\alpha^2/(\varepsilon_0\rho)]/(\omega+i0)^2$ for large $\omega$ in the
upper half-plane.

Finally, the contribution of ${\bf Q}_{\omega}$ is determined by the
integral
\begin{equation}
\int_{-\infty}^{\infty}d\omega\, \omega^3\, \bar{\bfsfG}({\bf r},{\bf
r}',-i\omega+0)\bar{\chi}({\bf
r}',-i\omega+0)\frac{1}{{\omega'}^2-(\omega+i0)^2}.
\label{A.6}
\end{equation}
As the integrand is proportional to $(\omega+i0)^{-3}$ for large $\omega$
in the upper half-plane, a contour deformation leads to a vanishing result.

From the above we conclude that the expression (\ref{A.2}) is equal to
\begin{equation}
-\frac{1}{\varepsilon_0}\, {\bm \Pi}({\bf r},0)
+\frac{1}{\varepsilon_0}\,[\alpha{\bf X}({\bf r},0)]_{\rm L} ,\label{A.7}
\end{equation}
which is in agreement with (\ref{3.13}).\\

\section{Commutators of the noise-current density\label{appendixb}}
In this appendix we derive the commutators of the noise-current density
with itself and with its hermitian conjugate. We start from (\ref{4.12}),
in which ${\bf J}({\bf r},\omega)$ is given as a linear combination of the
source terms $\bar{\bf J}({\bf r},p)$ and $\breve{\bf J}({\bf r},p)$. From
the definition (\ref{3.15}) and the canonical commutation relations
(\ref{2.4}) we obtain the commutator $[\bar{\bf J},\bar{\bf J}^{\dagger}]$ as
\widetext
\begin{equation}
\left[\bar{\bf J}({\bf r},p),[\bar{\bf J}({\bf r}',p')]^{\dagger}\right]
=\frac{i\hbarit}{\mu_0}\, \frac{p-{p'}^{\ast}}{p{p'}^{\ast}}\, 
\left({\bm \nabla}{\bm \nabla}-\bfsfI\,\Delta\right) \, 
\delta({\bf r}-{\bf r}')
-i\varepsilon_0\hbarit\, \frac{p{p'}^{\ast}}{p+{p'}^{\ast}}\,
\left[\bar{\chi}({\bf r},p)-\bar{\chi}({\bf r},{p'}^{\ast})\right]\,\bfsfI\,  
\delta({\bf r}-{\bf r}'). \label{B.1}
\end{equation}
Here we used the auxiliary relation
\begin{equation}
\int_0^{\infty} 
d\omega \, \frac{\omega^2 v^2_{\omega}}{(p^2+\omega^2)({p'}^2+\omega^2)}
=-\rho^2+\frac{\alpha^2\rho}{\varepsilon_0}\,
\frac{1}{p^2-{p'}^2}
\left[\frac{1}{\bar{\chi}({\bf r},p)}-\frac{1}{\bar{\chi}({\bf r},p')}\right],
\label{B.2}
\end{equation}
where for brevity we did not write the position dependence of $v_{\omega}$,
$\rho$ and $\alpha$. The commutator $[\breve{\bf J}({\bf r},p),[\breve{\bf
J}({\bf r}',p')]^{\dagger}]$ is equal to (\ref{B.1}), apart from an overall
minus sign.  Finally, the commutation relation of $\bar{\bf J}$ with
$\breve{\bf J}^{\dagger}$ is found to be
\begin{equation}
\left[\bar{\bf J}({\bf r},p),[\breve{\bf J}({\bf r}',p')]^{\dagger}\right]
=\frac{i\hbarit}{\mu_0}\frac{p+{p'}^{\ast}}{p{p'}^{\ast}}\, 
\left({\bm \nabla}{\bm \nabla}-\bfsfI\,\Delta\right) \, 
\delta({\bf r}-{\bf r}')
-i\varepsilon_0\hbarit\, \frac{p{p'}^{\ast}}{p-{p'}^{\ast}}\,
\left[\bar{\chi}({\bf r},p)-\bar{\chi}({\bf r},{p'}^{\ast})\right]\,\bfsfI\,  
\delta({\bf r}-{\bf r}'). \label{B.3}
\end{equation}
The last term at the right-hand side is not singular for
$p={p'}^{\ast}$. In fact, it is proportional to the derivative of the
susceptibility in that case. It should be noted that the commutators
(\ref{B.1}) and (\ref{B.3}) are local, as they vanish for ${\bf r}\neq {\bf
r}'$. This is not self-evident, since the definitions (\ref{3.15}) and
(\ref{3.21}) contain nonlocal longitudinal terms. When evaluating the
commutators, one finds that the nonlocal transverse delta function in the
canonical commutator (\ref{2.4a}) compensates for the nonlocal terms in
$\bar{\bf J}$ and $\breve{\bf J}$.

Having derived the commutators for $\bar{\bf J}$ and $\breve{\bf J}$, we
can determine the commutator of ${\bf J}({\bf r},\omega)$ with $[{\bf
J}({\bf r}',\omega')]^{\dagger}$, by using (\ref{4.12}). In $\bar{\bf J}$ we
have to substitute $p=-i\omega+0$ or $p'=-i\omega'+0$, while in $\breve{\bf
J}$ the argument is $p=i\omega+0$ or $p'=i\omega'+0$. Sorting out the
various terms, we get three different types of contributions, namely those
containing either no Green-function factor, or one or two such factors. In
the first contribution the terms proportional to $\left({\bm \nabla}{\bm
\nabla}-\bfsfI\,\Delta\right)$ drop out. The terms proportional to $\bfsfI\,
\delta({\bf r}-{\bf r}')$ cancel as well, unless $\omega$ and $\omega'$ are
equal:
\begin{equation}
\frac{1}{4\pi^2}
\left[ \bar{\bf J} ({\bf r},-i\omega+0)+\breve{\bf J} ({\bf r},i\omega+0)
, \left[\bar{\bf J}({\bf r}',-i\omega'+0)\right]^{\dagger}
+\left[\breve{\bf J}({\bf  r}',i\omega'+0)\right]^{\dagger}\right]=
\frac{\varepsilon_0\hbarit}{\pi}\, \omega^2\, \chi_i({\bf r},\omega)\, 
\delta(\omega-\omega')\, \bfsfI\, \delta({\bf r}-{\bf r}'). \label{B.4}
\end{equation}
The contribution with a single Green-function factor is:
\begin{equation}
\frac{\varepsilon_0\hbarit}{\pi^2 c^2}\, 
\frac{\omega\omega'}{\omega-\omega'-i0}\, 
\chi_i({\bf r},\omega)\chi_i({\bf r}',\omega')
\left[ \omega^2\, \bar{\bfsfG}({\bf r},{\bf r}',i\omega+0)-
{\omega'}^2\, \bar{\bfsfG}({\bf r},{\bf r}',-i\omega'+0)\right].
\label{B.5}
\end{equation}
Finally, the contribution with two Green-function factors is found as
\begin{eqnarray}
\frac{\varepsilon_0\hbarit}{\pi^2 c^2}\, \omega\omega'(\omega+\omega')
\chi_i({\bf r},\omega)\chi_i({\bf r}',\omega')\,
\int d{\bf r}''\, \bar{\bfsfG}({\bf r},{\bf r}'',i\omega+0)\cdot
\left\{{\bm \nabla}''\times\left[{\bm \nabla}''\times
\bar{\bfsfG}({\bf r}'',{\bf r}',-i\omega'+0)\right]\right\}&&\nonumber\\
&&\rule{-15.5cm}{0cm}+\frac{\varepsilon_0\hbarit}{\pi^2 c^4}\, 
\frac{\omega^3{\omega'}^3}{\omega-\omega'-i0}\, 
\chi_i({\bf r},\omega)\chi_i({\bf r}',\omega')\,
\int d{\bf r}''\, \left[\bar{\chi}({\bf r}'',i\omega+0)-
\bar{\chi}({\bf r}'',-i\omega'+0)\right]\,
\bar{\bfsfG}({\bf r},{\bf r}'',i\omega+0)\cdot
\bar{\bfsfG}({\bf r}'',{\bf r}',-i\omega'+0).
\label{B.6}
\end{eqnarray}
The second integral appearing here can be split into two parts, which may be
rewritten with the use of (\ref{4.1}) and (\ref{4.3}). After a partial
integration  we arrive at the identity
\begin{eqnarray}
\int d{\bf r}''\, \left[\bar{\chi}({\bf r}'',i\omega+0)-
\bar{\chi}({\bf r}'',-i\omega'+0)\right]\,
\bar{\bfsfG}({\bf r},{\bf r}'',i\omega+0)\cdot
\bar{\bfsfG}({\bf r}'',{\bf r}',-i\omega'+0)&&\nonumber\\
&&\rule{-10cm}{0cm}= -\frac{c^2}{\omega^2{\omega'}^2}\biggl[
\omega^2\, \bar{\bfsfG}({\bf r},{\bf r}',i\omega+0)
-{\omega'}^2\, \bar{\bfsfG}({\bf r},{\bf r}',-i\omega'+0)\nonumber\\
&&\rule{-8cm}{0cm}+(\omega^2-{\omega'}^2)\, 
\int d{\bf r}''\, \bar{\bfsfG}({\bf r},{\bf r}'',i\omega+0)\cdot
\left\{{\bm \nabla}''\times\left[
{\bm \nabla}''\times
\bar{\bfsfG}({\bf r}'',{\bf r}',-i\omega'+0)\right]\right\}\biggr].
\label{B.7}
\end{eqnarray}
For arbitrary $\omega$ and $\omega'$ this identity has the form of a
generalized optical theorem for the Green function. By putting
$\omega=\omega'$ one recovers the optical theorem that has been discussed
before \cite{DKW98}. When (\ref{B.7}) is used in (\ref{B.6}), it turns out
that all terms with spatial derivatives cancel, while the remaining terms
are the opposite of (\ref{B.5}). As a consequence, we are left with
(\ref{B.4}), so that we have proven the commutation relation (\ref{6.10}).

The commutator of ${\bf J}({\bf r},\omega)$ with ${\bf J}({\bf
r}',\omega')$ can be evaluated in a similar way. As a preparation, one
needs the commutators of $\bar{\bf J}$ and $\breve{\bf J}$ with their
counterparts for different arguments. These operators satisfy the
identities
\begin{equation}
[\bar{\bf J}({\bf r},p)]^{\dagger}=\bar{\bf J}({\bf r},p^{\ast})\quad , \quad 
[\breve{\bf J}({\bf r},p)]^{\dagger}=\breve{\bf J}({\bf r},p^{\ast}),
\label{B.8}
\end{equation}
as follows from inspection of (\ref{3.15}) and (\ref{3.21}). Hence, the
commutators of $\bar{\bf J}$ and $\breve{\bf J}$ can be written down
immediately by using (\ref{B.1}) and (\ref{B.3}). Subsequently, these
commutation relations can be employed in evaluating the commutator of ${\bf
J}({\bf r},\omega)$ with ${\bf J}({\bf r}',\omega')$. As before, we
encounter terms with various numbers of Green-function factors. The
contribution without a Green function is found to vanish:
\begin{equation}
\left[ \bar{\bf J} ({\bf r},-i\omega+0)+\breve{\bf J} ({\bf r},i\omega+0)
, \bar{\bf J}({\bf r}',-i\omega'+0)+\breve{\bf J}({\bf
  r}',i\omega'+0)\right]=0. \label{B.9}
\end{equation}
In an analogous way as above, the terms with one and with two
Green-function factors in the commutator $[{\bf J}({\bf r},\omega),{\bf
J}({\bf r}',\omega')]$ can be shown to cancel. This completes the proof of
(\ref{6.11}).\\

\section{Evaluation of the Hamiltonian\label{appendixc}}
In order to show how the Hamiltonian (\ref{2.3}) is recovered from
(\ref{6.16}), we give a few examples of the calculations that are
involved. We shall demonstrate how the terms quadratic in ${\bm \Pi}$ and
in ${\bf A}$ in (\ref{2.3}) are obtained after substitution of (\ref{5.3})
with (\ref{5.4a})-(\ref{5.4b}) in (\ref{6.16}). In this appendix all
canonical variables are taken at $t=0$.

We start with the term quadratic in ${\bm \Pi}$. After substitution of the
appropriate expressions we get the following contribution to the Hamiltonian:
\begin{equation}
\frac{1}{\pi\epsilon_0 c^4}\int d{\bf r}\int_0^{\infty} d\omega\, \omega^3\, 
\chi_i({\bf r},\omega)\int d{\bf r}'\int d{\bf r}''\, {\bm \Pi}({\bf r}')\cdot
\bar{\bfsfG}({\bf r}',{\bf r},-i\omega+0)\cdot
\bar{\bfsfG}({\bf r},{\bf r}'',i\omega+0)\cdot{\bm \Pi}({\bf r}''). \label{C.1}
\end{equation}
We could replace the transverse parts of the Green functions by the full
Green functions, as ${\bm \Pi}$ is purely transverse. Let us now rewrite
$\chi_i({\bf r},\omega)$ as $-(i/2)[\bar{\chi}({\bf r},-i\omega+0)-
\bar{\chi}({\bf r},i\omega+0)]$ . Subsequently, we carry out the integral
over ${\bf r}$ by means of the optical theorem, which follows from
(\ref{B.7}) by taking $\omega'=\omega$. In doing so, the integral with the
spatial derivatives in (\ref{B.7}) drops out, whereas the contributions
with a single Green function in (\ref{B.7}) remain. In this way we find
from (\ref{C.1}):
\begin{equation}
-\frac{i}{2\pi\varepsilon_0 c^2}\int_0^{\infty} d{\omega}\, \omega 
\int d{\bf r}'\int d{\bf r}''\, {\bm \Pi}({\bf r}')\cdot
\left[\bar{\bfsfG}({\bf r}',{\bf r}'',i\omega+0)-
\bar{\bfsfG}({\bf r}',{\bf r}'',-i\omega+0)\right]\cdot{\bm \Pi}({\bf
r}''). \label{C.2}
\end{equation}
The integral over the frequency can now be performed with the use of the sum
rule (\ref{A.4}). We finally obtain the simple result 
$(2\varepsilon_0)^{-1} \int d{\bf r}\, [{\bm \Pi}({\bf r})]^2$, as in
(\ref{2.3}).

As a second example, we consider the terms quadratic in ${\bf A}$. From
(\ref{6.16}) with (\ref{5.3}) and (\ref{5.4a}) we get a similar expression
as in (\ref{C.1}). The main difference is a factor $\omega^5$ instead of
$\omega^3$ in the integrand. After performing the integral over ${\bf r}$
as before, we arrive at
\begin{equation}
-\frac{i\varepsilon_0}{2\pi c^2}\int_0^{\infty} d{\omega}\, \omega^3 
\int d{\bf r}'\int d{\bf r}''\, {\bf A}({\bf r}')\cdot
\left[\bar{\bfsfG}({\bf r}',{\bf r}'',i\omega+0)-
\bar{\bfsfG}({\bf r}',{\bf r}'',-i\omega+0)\right]\cdot{\bf A}({\bf
r}''). \label{C.3}
\end{equation}
Owing to the presence of the factor $\omega^3$ we cannot use the sum rule
(\ref{A.4}). However, we may proceed by adding and subtracting the asymptotic
form (\ref{A.3}). In this way, the $\omega$-integral becomes : 
\begin{equation}
-\int_{-\infty}^{\infty}
 d{\omega}\, \omega^3\left[\bar{\bfsfG}({\bf r},{\bf r}',-i\omega+0)-
\frac{c^2}{(\omega+i0)^2}\, \bfsfI\, \delta({\bf r}-{\bf r}')\right],
\label{C.4}
\end{equation}
where we relabeled the position variables. Since the asymptotic form of the
susceptibility for large $\omega$ is
$-[\alpha^2/(\varepsilon_0\rho)]/(\omega+i0)^2$, as we have seen in
appendix \ref{appendixa}, one derives from the differential equation for
the Green function the asymptotic form for large $\omega$:
\begin{equation}
\bar{\bfsfG}({\bf r},{\bf r}',-i\omega+0)
-\frac{c^2}{(\omega+i0)^2}\, \bfsfI\, \delta({\bf r}-{\bf r}')
\simeq 
\frac{c^4}{(\omega+i0)^4}\left[ 
\left({\bm \nabla}{\bm \nabla}-\bfsfI\, \Delta\right)\, 
\delta({\bf r}-{\bf r}')+\frac{\mu_0\alpha^2}{\rho}
\, \bfsfI\, \delta({\bf r}-{\bf r}')\right], \label{C.5}
\end{equation}
as a generalization of (\ref{A.3}). Employing this asymptotic form in
(\ref{C.4}) one finds by contour integration the sum rule:
\begin{equation}
\int_{-\infty}^{\infty}
 d{\omega}\, \omega^3\left[\bar{\bfsfG}({\bf r},{\bf r}',-i\omega+0)-
\frac{c^2}{(\omega+i0)^2}\, \bfsfI\, \delta({\bf r}-{\bf r}')\right]=
-i\pi c^4 \left[ \left({\bm \nabla}{\bm \nabla}-\bfsfI\, \Delta\right)\, 
\delta({\bf r}-{\bf r}')+\frac{\mu_0 \alpha^2}{\rho}\,
\bfsfI\, \delta({\bf r}-{\bf r}')\right]. \label{C.6}
\end{equation}
Substitution in (\ref{C.3}) yields
\begin{equation}
\frac{1}{2\mu_0}\int d{\bf r}\, {\bf A}({\bf r})\cdot
\left({\bm \nabla}{\bm \nabla}-\bfsfI\, \Delta\right)\cdot{\bf A}({\bf r}) +
\int d{\bf r}\, \frac{\alpha^2}{2\rho}\, [{\bf A}({\bf r})]^2=
\frac{1}{2\mu_0}\int d{\bf r}\,[{\bm \nabla}\times {\bf A}({\bf r})]^2
+\int d{\bf r}\, \frac{\alpha^2}{2\rho}\, [{\bf A}({\bf r})]^2, \label{C.7}
\end{equation}
which agrees with the contributions in (\ref{2.3}). 

Similar techniques can be used to obtain the other terms in the Hamiltonian
(\ref{2.3}). Several additional sum rules, with integrands containing
products of the Green function and the susceptibility, are needed
in establishing complete agreement. These are:
\begin{subequations}
\label{C.8}
\begin{eqnarray}
&&\int_{-\infty}^{\infty} d{\omega}\, \omega^3\, 
\bar{\chi}({\bf r},-i\omega+0)\,
\bar{\bfsfG}({\bf r},{\bf r}',-i\omega+0)\, \bar{\chi}({\bf r}',-i\omega+0)=0,
\label{C.8a}\\
&&\int_{-\infty}^{\infty} d{\omega}\, \omega^5\, 
\bar{\chi}({\bf r},-i\omega+0)\,
\bar{\bfsfG}({\bf r},{\bf r}',-i\omega+0)\, \bar{\chi}({\bf r}',-i\omega+0)=
-\frac{i\pi c^2\alpha^4}{\varepsilon_0^2\rho^2}\, \bfsfI\, 
\delta({\bf r}-{\bf r}'). \label{C.8b}
\end{eqnarray}
\end{subequations}
Furthermore, one needs two sum rules for the susceptibility:
\begin{subequations}
\label{C.9}
\begin{eqnarray}
&&\int_{-\infty}^{\infty} d{\omega}\, \omega \, 
\bar{\chi}({\bf r},-i\omega+0)=
\frac{i\pi \alpha^2}{\varepsilon_0\rho},\label{C.9a}\\
&&\int_{-\infty}^{\infty} d{\omega}\, \omega^3 \, \left[
\bar{\chi}({\bf r},-i\omega+0)+\frac{\alpha^2}{\varepsilon_0 \rho}\, 
\frac{1}{(\omega+i0)^2}
\right]= \frac{i\pi\alpha^2 \tilde{\omega}_0^2}{\varepsilon_0\rho}.\label{C.9b}
\end{eqnarray}
\end{subequations}
 To prove these sum rules one uses a contour deformation and the asymptotic
behavior of the integrands, as before. \\

\section{Time-dependent operators\label{appendixd}}

In this appendix we show how time-dependent operators can be expressed as
linear combinations of the canonical variables. As examples we shall
discuss the vector potential, the electric field and the polarization
density.

The time-dependent vector potential has been given in (\ref{6.17}). By
substituting the expression (\ref{5.3}) for the noise-current density, and
employing the same methods as used in checking (\ref{6.14}) we derive
\begin{eqnarray} 
{\bf A}({\bf r},t)&=&\int d{\bf r}' \int_0^{\infty}d\omega \, e^{-i\omega t}\, 
\biggl\{  
\bfsfc_{AA}({\bf r},{\bf r}',\omega)\cdot{\bf A}({\bf r}',0)
+\bfsfc_{A\Pi}({\bf r},{\bf r}',\omega)\cdot{\bm \Pi}({\bf r}',0)
+\bfsfc_{AX}({\bf r},{\bf r}',\omega)\cdot{\bf X}({\bf r}',0)\nonumber\\
&&\rule{-2cm}{0cm} +\bfsfc_{AP}({\bf r},{\bf r}',\omega)\cdot{\bf P}({\bf r}',0)
+\int_0^{\infty}d\omega'\, 
\bfsfc_{AYQ}({\bf r},{\bf r}',\omega,\omega')\cdot
\left[{\omega'}^2\, {\bf Y}_{\omega'}({\bf r}',0)+
\frac{i\omega}{\rho'}\, 
{\bf Q}_{\omega'}({\bf r}',0)\right] \biggr\} + \text{h.c.}\; . 
\label{D.1}
\end{eqnarray} 
The coefficients are
\begin{subequations}
\label{D.2}
\begin{eqnarray}
\bfsfc_{AA}({\bf r},{\bf r}',\omega)
&=&-\frac{1}{\pi c^2}\, \omega \, {\rm Im}
\left[\bfsfG_{{\rm TT}'}({\bf r},{\bf r}',\omega)\right],\label{D.2a}\\
\bfsfc_{A\Pi}({\bf r},{\bf r}',\omega)
&=&-\frac{i\mu_0}{\pi}\, {\rm Im}
\left[\bfsfG_{{\rm TT}'}({\bf r},{\bf r}',\omega)\right] ,\label{D.2b}\\
\bfsfc_{AX}({\bf r},{\bf r}',\omega)
&=&
-\frac{i\mu_0\alpha'}{\pi}\,  {\rm Im}
\left[\bfsfG_{{\rm TT}'}({\bf r},{\bf r}',\omega)\right]
-\frac{i\rho'}{\pi c^2\alpha'}\, \omega^2\, 
{\rm Im}
\left[\bfsfG_{\rm T}({\bf r},{\bf r}',\omega)\, \chi({\bf r}',\omega)\right] 
,\label{D.2c}\\
\bfsfc_{AP}({\bf r},{\bf r}',\omega)
&=&\frac{1}{\pi c^2\alpha'}\,  \omega\, 
{\rm Im}
\left[\bfsfG_{\rm T}({\bf r},{\bf r}',\omega)\, \chi({\bf r}',\omega)\right],
\label{D.2d}\\
\bfsfc_{AYQ}({\bf r},{\bf r}',\omega,\omega')
&=&\frac{1}{\pi c^2\alpha'}\,  
{\rm Im}\left[
\frac{\omega\, v'(\omega')}{{\omega'}^2-(\omega +i0)^2}\,
\bfsfG_{\rm T}({\bf r},{\bf r}',\omega)\, \chi({\bf r}',\omega)\right].
\label{D.2e}
\end{eqnarray}
\end{subequations}

The expression (\ref{D.1}) gives the vector potential ${\bf A}({\bf r},t)$
for all $t$ in terms of the canonical variables at $t=0$. In particular, it
may be used to determine the vector potential for large $t$, when all
transients have died out. In order to derive this asymptotic form, one
starts by noting that the term with the hermitian conjugate in (\ref{D.1})
can be used to extend the $\omega$-integral over the whole real axis. For
all positive $t$ this integral may be evaluated by deforming the contour in
the lower half-plane. If $t$ gets large, the behavior of ${\bf A}({\bf
r},t)$ is dominated by the contributions from those singularities of the
frequency-dependent integrand that are located close to the real
$\omega$-axis in the lower half-plane. To find these singularities we
consider the contributions from the various terms in (\ref{D.1}) one by
one. The contributions involving ${\bf A}$, ${\bm \Pi}$, ${\bf X}$ and
${\bf P}$ depend on the coefficients (\ref{D.2a})-(\ref{D.2d}). The
singularities in these coefficients are determined by those of (the
analytical continuations of) $\chi({\bf r},\omega)$ and $\bfsfG({\bf
r},{\bf r}',\omega)$. When the susceptibility has a finite imaginary part
for real $\omega$, continuity implies that the singularities of $\chi$ and
$\bfsfG$ in the lower half-plane are located at a finite distance from the
real $\omega$-axis. As a consequence, the contributions from the terms with
${\bf A}$, ${\bm \Pi}$, ${\bf X}$ and ${\bf P}$ in (\ref{D.1}) will die out
exponentially fast for large $t$. On the other hand, some of the
singularities of the coefficient (\ref{D.2e}) are really close to the real
axis, as they are given by $\omega=\pm\omega'-i0$. The contributions from
these singularities will dominate the behavior of (\ref{D.1}) for large
$t$. These contributions are readily evaluated by calculating the residues.
One arrives at the result given in (\ref{6.18}) of the main text. It shows
that the long-time behavior of ${\bf A}({\bf r},t)$ is governed by the
specific combination ${\bf J}_l({\bf r},\omega)$ of bath operators, given
in (\ref{6.19}):
\begin{equation}
{\bf A}({\bf r},t)\simeq -\mu_0\int d{\bf r}'\int_0^{\infty}d\omega\, 
e^{-i\omega t}\, 
\bfsfG_{\rm T}({\bf r},{\bf r}',\omega)\cdot
{\bf J}_l({\bf r}',\omega) + {\rm h.c.}\; .
\label{D.3}
\end{equation}
As noted in the main text, the combination of bath operators occurring in
${\bf J}_l({\bf r},\omega)$ is in fact proportional to the annihilation
operator of the bath harmonic oscillator at ${\bf r}$ and with frequency
$\omega$. 

The time-dependent electric field ${\bf E}({\bf r},t)$ can be evaluated in
an analogous way. Substituting (\ref{5.3}) in (\ref{6.20}) we obtain an
expression like (\ref{D.1}). The coefficients $\bfsfc_{Ei}$ (with
$i=A,\Pi,X,P,YQ$) follow from $\bfsfc_{Ai}$ in (\ref{D.2}) upon multiplying
each coefficient by $i\omega$ and dropping the subscript ${\rm T}$ (but not
${\rm T}'$) of the Green functions.  The analysis of the long-time behavior
of the electric field is completely analogous to that of the vector
potential. One finds an expression of the same form as (\ref{6.20}), with
${\bf J}$ replaced by ${\bf J}_l$:
\begin{equation}
{\bf E}({\bf r},t)\simeq -i\mu_0 \int d{\bf r}' \int_0^{\infty}
d \omega\, e^{-i\omega t}\,\omega \, \bfsfG({\bf r},{\bf r}',\omega)
\cdot{\bf J}_l({\bf r}',\omega) +\text{h.c.}\; . \label{D.4}
\end{equation}

Finally, we consider the time-dependent polarization density. From
(\ref{6.21}) and (\ref{5.3}) with (\ref{5.4}) we derive an expression for
${\bf X}({\bf r},t)$, which has the same form as (\ref{D.1}), with the
coefficients:
\begin{subequations}
\label{D.5}
\begin{eqnarray}
\bfsfc_{XA}({\bf r},{\bf r}',\omega)
&=&\frac{i\varepsilon_0}{\pi c^2\alpha}\, \omega^2 \, {\rm Im}
\left[\chi({\bf r},\omega)\, 
\bfsfG_{{\rm T}'}({\bf r},{\bf r}',\omega)\right],\label{D.5a}\\
\bfsfc_{X\Pi}({\bf r},{\bf r}',\omega)
&=&-\frac{1}{\pi c^2\alpha}\, \omega\,{\rm Im}
\left[\chi({\bf r},\omega)\,
\bfsfG_{{\rm T}'}({\bf r},{\bf r}',\omega)\right] ,\label{D.5b}\\
\bfsfc_{XX}({\bf r},{\bf r}',\omega)
&=&\frac{\varepsilon_0\rho}{\pi \alpha^2}\, \omega\,
\chi_i({\bf r},\omega)\,
\bfsfI\, \delta({\bf r}-{\bf r}')
-\frac{\alpha'}{\pi c^2\alpha}\, \omega\, 
{\rm Im}
\left[\chi({\bf r},\omega)\,
\bfsfG_{{\rm T}'}({\bf r},{\bf r}',\omega)\, \right] \nonumber\\
&&-\frac{\varepsilon_0\rho'}{\pi c^2\alpha\alpha'}\, \omega^3  {\rm Im}
\left[\chi({\bf r},\omega)\,
\bfsfG({\bf r},{\bf r}',\omega)\chi({\bf r}',\omega)\right] 
,\label{D.5c}\\
\bfsfc_{XP}({\bf r},{\bf r}',\omega)
&=&\frac{i\varepsilon_0}{\pi\alpha^2}\, \chi_i({\bf r},\omega)\, 
\bfsfI\, \delta({\bf r}-{\bf r}')
-\frac{i\varepsilon_0}{\pi c^2\alpha\alpha'}\,  \omega^2 \, 
{\rm Im}\left[\chi({\bf r},\omega)\,
\bfsfG({\bf r},{\bf r}',\omega)\, \chi({\bf r}',\omega)\right],
\label{D.5d}\\
\bfsfc_{XYQ}({\bf r},{\bf r}',\omega,\omega')
&=&\frac{i\varepsilon_0}{\pi\alpha^2}\, 
{\rm Im}\left[\frac{v(\omega')}{{\omega'}^2-(\omega+i0)^2}\,
\chi({\bf r},\omega)\right]\, \bfsfI\, \delta({\bf r}-{\bf r}')\nonumber\\
&&-\frac{i\varepsilon_0}{\pi c^2 \alpha\alpha'}\,  
{\rm Im}\left[
\frac{\omega^2\, v'(\omega')}{{\omega'}^2-(\omega +i0)^2}\,
\chi({\bf r},\omega)\,
\bfsfG({\bf r},{\bf r}',\omega)\, \chi({\bf r}',\omega)\right].
\label{D.5e}\
\end{eqnarray}
\end{subequations}
As before, the behavior in the long-time limit is dominated by the
singularities close to the real frequency axis. These singularities arise
from the coefficient in (\ref{D.5e}). Evaluating their contributions we
arrive at the following expression for ${\bf X}({\bf r},t)$ in the
long-time limit:
\begin{equation}
{\bf X}({\bf r},t)\simeq\frac{i}{c^2\alpha}
\int d{\bf r}'\int_0^{\infty}d\omega\, 
e^{-i\omega t}\, \omega \, \chi({\bf r},\omega)\, 
\bfsfG({\bf r},{\bf r}',\omega)\cdot{\bf J}_l({\bf r}',\omega)
-\frac{i}{\alpha}\int_0^{\infty} d\omega\, e^{-i\omega t}\, \frac{1}{\omega}\,
{\bf J}_l({\bf r},\omega)+\text{h.c.}\; , \label{D.6}
\end{equation}
with ${\bf J}_l$ given by (\ref{6.19}). Upon multiplying with $-\alpha$ and
comparing with (\ref{6.22}), we see that the long-time limit has the same
effect on the polarization density as it has on the vector potential and
the electric field: the dependence on the full noise-current density ${\bf
J}({\bf r},\omega)$ is replaced by a dependence on ${\bf J}_l({\bf
r},\omega)$.

\twocolumngrid

\end{document}